\documentclass[prd,nofootinbib,showpacs,superscriptaddress, preprint]{revtex4-1}

\usepackage[T1]{fontenc}
\usepackage{amsmath,amssymb}
\usepackage{epsfig}
\usepackage{graphicx}
\usepackage[usenames,dvipsnames]{color}
\usepackage{slashed}
\usepackage[colorlinks,citecolor=blue]{hyperref}
\usepackage{color}
\begin{document} 
\title{Neutrinoless double beta decay in minimal left-right symmetric model with universal seesaw}


\author{Debasish Borah}
\email{dborah@iitg.ac.in}
\affiliation{Department of Physics, Indian Institute of Technology Guwahati, Assam 781039, India}
\author{Arnab Dasgupta}
\email{arnabdasgupta@protonmail.ch}
\affiliation{%
 School of Liberal Arts, Seoul-Tech, Seoul 139-743, Korea}
\affiliation{Theoretical Physics Division, Physical Research Laboratory, Navrangpura, Ahmedabad, Gujarat 380009, India}
\author{Sudhanwa Patra}
\email{sudhanwa@iitbhilai.ac.in}
\affiliation{Department of Physics, Indian Institute of Technology Bhilai, Govt. Engg. College Campus, Sejbahar, Raipur, Chhattisgarh 492015, India}
\affiliation{Center of Excellence in Theoretical and Mathematical Sciences, 
Siksha `O' Anusandhan University, Bhubaneswar 751030, India}

\begin{abstract}
We present a detailed discussion on neutrinoless double beta decay $(0\nu \beta \beta)$ within left-right symmetric models based on the gauge symmetry of type $SU(2)_L \times SU(2)_R \times U(1)_{B-L}$ as well as $SU(3)_L \times SU(3)_R \times U(1)_{X}$ where
fermion masses including that of neutrinos are generated through a universal seesaw mechanism. We find that one or more of the right-handed neutrinos could be as light as a few keV if left-right symmetry breaking occurs in the range of a few TeV to 100 TeV. With such light right-handed neutrinos, we perform a detailed study of new physics contributions to $0\nu \beta \beta$ and constrain the model parameters from the latest experimental bound on such a rare decay process. We find that the new physics contribution to $0\nu \beta \beta$ in such a scenario, particularly the heavy-light neutrino mixing diagrams, can individually saturate the existing experimental bounds, but their contributions to total $0\nu \beta \beta$ half-life cancels each other due to unitarity of the total $6\times 6$ mass matrix. The effective contribution to half-life therefore, arises from the purely left and purely right neutrino and gauge boson mediated diagrams.
We find that the parameter space saturating the $0\nu \beta \beta$ bounds remain allowed from the latest experimental bounds on charged lepton flavour violating decays like $\mu \rightarrow e \gamma$. We finally include the bounds from cosmology and supernova to constrain the parameter space of the model.
\end{abstract}

\maketitle
%
\section{Introduction} 
\label{sec:intro}
The Standard Model (SM) of particle physics has been established as the most successful description of the fundamental particles and their interactions: strong, weak and electromagnetic. The model based on the local gauge symmetry $SU(2)_L \times U(1)_Y \times SU(3)_C$ 
describing strong, weak and electromagnetic interactions between fundamental particles gets broken down to $U(1)_Q \times SU(3)_c$ remnant gauge symmetry spontaneously due to the non-zero vacuum expectation value (vev) of the Higgs field charged under the $SU(2)_L \times U(1)_Y$ symmetry of the model. Since the discovery of the Higgs boson in 2012 at the large hadron collider (LHC), the SM has been confirmed again and again as the only theory around the electroweak scale with no signs of new physics yet. In spite of these null results for new physics beyond the standard model (BSM), there are convincing amount of evidence suggesting the presence of new physics. This need for new physics arises due to the inadequacies of the SM as it can not address several observed phenomena as well as theoretical questions. Non zero but tiny neutrino mass \cite{Olive:2016xmw} is one such observation which the SM can not address. Due to the absence of right handed neutrinos in the SM, there is no renormalisable interaction 
between the neutrino and the Higgs field resulting in vanishing mass of neutrinos. The SM also can not explain the origin of parity violation seen in low energy 
weak interaction processes which is at sharp contrast with other interactions like electromagnetic and strong which are parity conserving. This motivates one to speculate that all fundamental interactions are parity conserving at the most fundamental level or at a very high energy scale and the SM is a parity violating low energy manifestation of such a unified parity conserving theory. These two observations namely, non-zero neutrino mass and parity violation in weak interactions can be explained naturally within the framework of the left-right symmetric model (LRSM)~\cite{Pati:1974yy, Mohapatra:1974gc, Senjanovic:1975rk}-based on the gauge symmetry $SU(2)_L \times SU(2)_R \times U(1)_{B-L} \times SU(3)_C$  where both left and right-handed fermions are treated on equal footing in a parity symmetric manner. The right handed fermions transform as doublets under the new $SU(2)_R$ gauge symmetry similar to the transformation of left handed fermions under the $SU(2)_L$ gauge symmetry of the SM. The inclusion of 
right handed neutrinos become a necessity in such a framework and hence neutrinos can naturally acquire a non-zero mass.

In the framework of the LRSM, the light neutrino masses can arise in several different ways depending on the scalar content and the way spontaneous symmetry breaking occurs from LRSM gauge symmetry to that of the SM and finally to the $U(1)_Q \times SU(3)_c$ symmetry. The minimal version of LRSM contains scalars which transform as $SU(2)$ triplets and bidoublet with the triplets playing to role of breaking the LRSM gauge symmetry to that of the SM and the bidoublet playing the same role in electroweak symmetry breaking. This promising theory of parity preserving weak interaction can have testable consequences for different experiments or observed phenomena when LRSM gauge symmetry breaking occurs at few TeV. Such tantalising consequences can be in the gauge sector in terms of additional gauge bosons ~\cite{Keung:1983uu,  Ferrari:2000sp,  Schmaltz:2010xr, Nemevsek:2011hz, Chen:2011hc, Chakrabortty:2012pp, Das:2012ii, 
AguilarSaavedra:2012gf, Han:2012vk, Chen:2013fna, Rizzo:2014xma,  Deppisch:2014zta, Deppisch:2015qwa, Gluza:2015goa, 
Ng:2015hba, Patra:2015bga,  Dobrescu:2015qna, Brehmer:2015cia, Dev:2015pga,  Coloma:2015una, Deppisch:2015cua, Dev:2015kca, 
Mondal:2015zba, Aguilar-Saavedra:2015iew,    Lindner:2016lpp, Lindner:2016lxq, Mitra:2016kov, Anamiati:2016uxp, 
Khachatryan:2014dka, Aad:2015xaa, Khachatryan:2016jqo}, 
in the Higgs sector~\cite{Gunion:1989in, Deshpande:1990ip, Polak:1991vf, Barenboim:2001vu, Azuelos:2004mwa, Jung:2008pz, 
Bambhaniya:2013wza, Dutta:2014dba, Bambhaniya:2014cia,   Bambhaniya:2015wna, Dev:2016dja, ATLAS:2014kca, CMS:2016cpz, ATLAS:2016pbt}, 
in the context of neutrinoless double beta decay~\cite{Mohapatra:1980yp, Mohapatra:1981pm, Picciotto:1982qe, 
Hirsch:1996qw, Arnold:2010tu, Tello:2010am,  Chakrabortty:2012mh, Nemevsek:2012iq, Patra:2012ur, Awasthi:2013ff, Barry:2013xxa, 
Dev:2013vxa, Huang:2013kma, Dev:2014xea, Ge:2015yqa, Borah:2015ufa,  Awasthi:2015ota,   Horoi:2015gdv, Bambhaniya:2015ipg, 
Gu:2015uek, Borah:2016iqd, Awasthi:2016kbk}, in dark matter contexts~\cite{Heeck:2015qra,Patra:2015qny,Borah:2016uoi,Borah:2016ees,
Borah:2017xgm,Patra:2015vmp}
in low-energy charged lepton flavour violation (LFV)~\cite{Riazuddin:1981hz, Pal:1983bf, Mohapatra:1992uu, Cirigliano:2004mv, 
Cirigliano:2004tc, Bajc:2009ft, Tello:2010am, Das:2012ii, Barry:2013xxa, Dev:2013oxa,  Borah:2013lva, Chakrabortty:2015zpm, 
Awasthi:2015ota, Bambhaniya:2015ipg, Borah:2016iqd, Lindner:2016bgg, Bonilla:2016fqd} 
and electric dipole moment (EDM)~\cite{Ecker:1983dj, Nieves:1986uk, Frere:1991jt, Nemevsek:2012iq, Dev:2014xea, Maiezza:2014ala}.  Apart from this most widely studied minimal LRSM, there have been alternative formulation of the left-right symmetric models as well which have different scalar content and different ways of generating fermion masses. For example, if the scalar sector of the minimal LRSM (MLRSM) is replaced by a pair of scalar doublets transforming under $SU(2)_L, SU(2)_R$ respectively then the desired symmetry breaking can be achieved in a more minimal way. However, the absence of the bidoublet prevents one from writing renormalisable mass terms for the fermions forcing one to introduce higher dimensional operators \cite{Brahmachari:2003wv}. A renormalisable version of such LRSM with universal seesaw for all fermions can be achieved by introducing additional heavy fermions \cite{Davidson:1987tr, Babu:1988mw, Gu:2010zv}. Several other realisations of fermion masses within such LRSM without scalar 
bidoublet can be found in \cite{Borah:2016lrl, Borah:2016hqn, Borah:2017leo}. The LRSM with universal seesaw (LRSM-US) for all fermions was also studied from $0\nu \beta \beta$ point of view in \cite{Gu:2010yf, Patra:2012ur, Deppisch:2017vne} and more recently within the $331$ set up \cite{Borah:2017inr}. The extension of such a framework to the $331$ models namely, the $SU(3)_L \times SU(3)_R \times U(1)_{X}$ models has also been studied recently \cite{Borah:2017inr}. In the $331$ version of LRSM with universal seesaw, it was pointed out that due to the existence of light right handed neutrinos in the keV-MeV range, such a model can give observable contributions to $0\nu \beta \beta$ due to large light-heavy neutrino mixing. Such a version of LRSM where the scalar sector can be made more minimal than MLRSM at the price of introducing extra vector-like iso-singlet quarks and leptons was of much interest in the context of LHC anomalies~\cite{Dev:2015vjd, Dasgupta:2015pbr, Deppisch:2016scs}. Apart from having all other generic features of minimal LRSM in terms of providing an explanation to the origin neutrino mass, origin of parity violation, strong CP problem, allowing the possibility of non-supersymmetric grand unification the LRSM-US can also explain the origin of fermion mass hierarchies through seesaw mechanism instead of arbitrarily fine tuning the Yukawa couplings.

In the present work, we intend to study the contribution of different particles in LRSM-US to lepton number violating rare decay process of neutrinoless double beta decay that have been looked for at several experiments resulting in strict upper bound (lower bound) on the amplitude (half-life) for such a process. This is an extension of the recent work \cite{Borah:2017inr} to do a complete scan of all parameters, though confined to $SU(2)_L \times SU(2)_R \times U(1)_{B-L}$ version of LRSM for simplicity instead of the $331$ version of \cite{Borah:2017inr}. With the present $0\nu \beta \beta$ experiments like KamLAND-Zen \cite{Gando:2012zm, KamLAND-Zen:2016pfg}, GERDA \cite{Agostini:2013mzu, Agostini:2017iyd} probing the quasi-degenerate regime of light neutrino masses, one can expect the next generation experiments to cover the entire parameter space for $0\nu \beta \beta$, at least in the case inverted hierarchical pattern of light neutrino masses. The current lower limit on the half-life of this rare process from these two experiments lie in the range of $10^{25}-10^{26}$ year. The projected sensitivity of the phase III of KamLAND-Zen is $T_{1/2} > 2\times 10^{26}$ year after two years of data taking. Similar goal is also set by the GERDA experiment to 
reach $T_{1/2} > 10^{26}$ year. We show that the contributions to $0\nu\beta\beta$ in LRSM-US can saturate these experimental bounds even if the scale of $SU(2)_R \times U(1)_{B-L}$ breaking is outside the reach of present collider experiments. In particular, such a model allows the right handed neutrinos to be as light as a keV without any fine-tuning even if the symmetry breaking scale is kept around a few (tens of) TeV. This also allows for the possibility of large light-heavy neutrino mixing and can contribute significantly to $0\nu \beta \beta$ through light-heavy or left-right mixing diagrams. However, their combined contribution gets cancelled out due to the unitarity of the full $6\times 6$ mass matrix leaving out the purely left and purely right handed contributions to the $0\nu\beta\beta$ half-life. On the other hand, such large light-heavy mixing between neutrinos can introduce non-unitary effects to leptonic mixing and can be constrained significantly from charged lepton flavour violating decay like $\mu 
\rightarrow e \gamma$ that have been looked for at ongoing experiments like MEG \cite{TheMEG:2016wtm}. We constrain the model parameters from the requirement of satisfying both $0\nu\beta\beta$ and LFV constraints from the latest experimental data.

We first discuss neutrinoless double beta decay in the framework of left-right symmetric model with universal seesaw in two regimes:
i) firstly considering the $SU(2)_R$ gauge boson masses $M_{W_R} \approx 3~$TeV and equivalently the right handed neutrino masses $M_R \approx \mathcal{O}(\mbox{keV})$, ii) secondly, with $M_{W_R} 
\approx 50~$TeV (or equivalently, $M_R \approx \mathcal{O}(\mbox{MeV})$). In the former case, the diagrams mediated by $W_L-W_R$ as well as $W_L-W_L$ where light-heavy neutrino mixing can play the major role. On the other hand, the purely $W_R$ mediated diagram remains suppressed in this case. Similarly, with $M_{W_R} \approx 50~$TeV, the $W_R-W_R$ mediated diagrams give negligible contribution to $0\nu\beta\beta$. The dominant contribution in the latter case arises 
from $W_L-W_R$ mediated diagrams where large light-heavy neutrino mixing plays an important role, once again. The $W_L-W_R$ mixing diagram (so called $\eta$ diagram) can also contribute, sizeably in this case, though remain suppressed from experimental bounds. The importance of light-heavy neutrino mixing in the study of $0\nu \beta \beta$ within generic LRSM with type I/II seesaw was pointed out by \cite{Dev:2014xea, Huang:2013kma}. Since all the neutrinos are light, having masses smaller than the typical momentum exchange of the $0\nu\beta\beta$ process, we find that the combined contribution of such heavy light mixing to the half-life cancels out due to unitarity of the mass matrix. However, both left and right handed neutrinos can contribute sizeably due to their individual gauge interactions, saturating the experimental bounds on $0\nu\beta\beta$ half-life. After showing the new physics contribution to $0\nu \beta \beta$ for some benchmark values of parameters, we also scan the entire parameter space and put the constraints on $W_R$ mass, light-heavy neutrino mixing parameter as well as the lightest neutrino mass from the requirement of satisfying the current experimental bounds on $0\nu \beta \beta$. We also check that for the entire region of our interest with $W_R$ mass being varied all the way upto 100 TeV, the gauge boson mediated diagrams contributing to $\mu \rightarrow e \gamma$ remain very much suppressed compared to the latest experimental bound \cite{TheMEG:2016wtm}. Apart from using the latest experimental constraints on light neutrino parameters that have appeared in global fit work \cite{Esteban:2016qun}, we also use the limit on the sum of light neutrino masses from the Planck mission data as $\sum_i m_i < 0.17~$eV \cite{Ade:2015xua}.

The paper is organised as follows. In section \ref{sec1}, we briefly discuss the LRSM with universal seesaw for all fermions. In section \ref{sec2} we discuss different contributions to $0\nu\beta\beta$ in the model followed by a brief discussion on possible new physics contribution to the charged lepton flavour violating decay $\mu \rightarrow e \gamma$ in section \ref{sec3}. We discuss our numerical calculations in section \ref{sec4}, followed by cosmology, supernova bounds in section \ref{sec4a} and finally conclude in section \ref{sec5}.

\section{Left-Right Symmetric Model with Universal Seesaw}
\label{sec1}
In this section, we briefly recapitulate the left-right symmetric model without scalar bidoublet 
where all fermion masses are generated by a common universal seesaw mechanism. The particle content of the model transforms non trivially under the gauge symmetry of the model given by
\begin{equation}
\mathcal{G}_{LR} \equiv SU(2)_L \times SU(2)_R \times U(1)_{B-L} \times SU(3)_c \,,
\end{equation}
which gets broken down to the $U(1)_Q \times SU(3)_c$ of electromagnetism and colour spontaneously at two stages such that the electromagnetic charge $Q$ is defined as
\begin{equation}
	Q = T_{3L} + T_{3R} + \frac{B-L}{2} = T_{3L} + Y \,.
\end{equation}
We denote  $T_{3L}$ and $T_{3R}$ are, respectively, the third component of isospin corresponding to 
the gauge groups $SU(2)_L$ and $SU(2)_R$, and $Y$ is the hypercharge. Here difference between 
baryon and lepton number is promoted to local gauge symmetry. It should be noted that the model also has an in built discrete $Z_2$ symmetry or left-right symmetry (D parity) under which forces the couplings in the left and right sectors equal, making the theory left-right symmetric. The fermion content of the model is 
\begin{gather}
	Q_{L}=\begin{pmatrix}u_{L}\\
	d_{L}\end{pmatrix}, \quad Q_{R}=\begin{pmatrix}u_{R}\\
	d_{R}\end{pmatrix}\,,\nonumber \\
	\ell_{L}=\begin{pmatrix}\nu_{L}\\
	e_{L}\end{pmatrix}, \quad 
	\ell_{R}=\begin{pmatrix}\nu_{R}\\
	e_{R}\end{pmatrix} \,, 
\end{gather}
plus additional vector-like quarks and charged leptons,
\begin{gather}
	U_{L,R}\,,\quad D_{L,R}\, \quad E_{L,R}\, , \quad N_{L,R}\, .
\end{gather}
The spontaneous symmetry breaking is implemented with a scalar sector consisting of $SU(2)_{L,R}$ doublets $H_L\oplus H_R$ and the conventional scalar bidoublet of MLRSM is absent. All these fields with their transformations under the gauge symmetry are shown in table \ref{tab:LR2}.
\begin{table}[h]
\begin{center}
\begin{tabular}{|c|c|c||c|c|}
\hline
Field     & $ SU(2)_L$ & $SU(2)_R$ & $U(1)_{B-L}$ & $SU(3)_c$ \\
\hline
$Q_L$     &  2         & 1         & 1/3   & 3   \\
$Q_R$     &  1         & 2         & 1/3   & 3   \\
$\ell_L$  &  2         & 1         & -1    & 1   \\
$\ell_R$  &  1         & 2         & -1    & 1   \\
\hline
$U_{L,R}$ &  1         & 1         & 4/3   & 3   \\
$D_{L,R}$ &  1         & 1         & -2/3  & 3   \\
$E_{L,R}$ &  1         & 1         & -2    & 1   \\
$N_{L,R}$ &  1         & 1         & 0     & 1   \\
\hline
 $H_L$    &  2         & 1         & -1    & 1   \\
 $H_R$    &  1         & 2         & -1    & 1   \\
\hline
\end{tabular}
\end{center}
\caption{Field content and their transformations under the gauge symmetry of LRSM with universal seesaw.}
\label{tab:LR2}
\end{table}
The scalar potential of the model can be written as
\begin{align}
V & =\mu^2_{H} \left( H^{\dagger}_L H_L + H^{\dagger}_R H_R \right)+\lambda_{H} 
\left( (H^{\dagger}_L H_L)^2 + (H^{\dagger}_R H_R)^2 \right) + \lambda^{\prime}_{H} (H^{\dagger}_L 
H_L) (H^{\dagger}_R H_R)
\end{align}
The scalar fields can acquire non-zero vev as
\begin{align}
	\langle H_R \rangle = \begin{pmatrix} \frac{v_R}{\sqrt{2}} \\ 0 \end{pmatrix}, \quad 
	\langle H_L \rangle = \begin{pmatrix} \frac{v_L}{\sqrt{2}} \\ 0 \end{pmatrix}. 
\end{align}
The vev of the neutral component of $H_R$ spontaneously breaks the symmetry of the LRSM to that of the SM while the vev of the neutral component of $H_L$ gives rise to the usual electroweak symmetry breaking. In other words, the desired symmetry breaking chain is
$$ SU(2)_L \times SU(2)_R \times U(1)_{B-L} \quad \underrightarrow{\langle
H^0_R \rangle} \quad SU(2)_L\times U(1)_Y \quad \underrightarrow{\langle H^0_L \rangle} \quad U(1)_{Q}$$
In the scalar potential written above, the discrete left-right symmetry is assumed which ensures the equality 
of left and right sector couplings. However, as shown in earlier works \cite{Brahmachari:2003wv, Davidson:1987tr, Babu:1988mw, Gu:2010zv} the scalar potential of such a model with exact discrete left-right 
symmetry is too restrictive and gives to either parity preserving $(v_L =v_R)$ solution or a solution with 
$(v_R \neq 0, v_L = 0)$ at tree level. While the first one is not phenomenologically acceptable the latter 
solution can be acceptable if a non-zero vev $v_L \neq 0$ can be generated through radiative corrections 
\cite{Kobakhidze:2013pya}. While it may naturally explain the smallness of $v_L$ compared to $v_R$, it will constrain 
the parameter space significantly \cite{Kobakhidze:2013pya}. Another way of achieving a parity breaking vacuum is to consider 
softly broken discrete left-right symmetry by considering different mass terms for the left and right sector 
scalars \cite{Mohapatra:1974gc, Brahmachari:2003wv, Davidson:1987tr, Babu:1988mw, Gu:2010zv}. As it was pointed out by the authors of \cite{Mohapatra:1974gc}, such a model which 
respects the discrete left-right symmetry everywhere except in the scalar mass terms, preserve the \textit{naturalness} 
of the left-right symmetry in spite of radiative corrections. Another interesting way is to achieve parity breaking 
vacuum is to decouple the scale of parity breaking and gauge symmetry breaking by introducing a parity odd singlet 
scalar \cite{Chang:1983fu}. In this work, we simply assume that the desired symmetry breaking can be achieved by considering different mass terms for left and right sector scalars \cite{Mohapatra:1974gc} without incorporating any new field content. Such a minimal assumption is not going to affect our discussion of $0\nu\beta\beta$ and LFV or even the origin of fermion masses.%

After the neutral components of the scalar fields acquire non-zero vev's, the resulting gauge boson masses can be derived as 
$$ M_{W_L} = \frac{g}{2}v_{L}, \;\; M_{W_R} =\frac{g}{2}v_{R}, \;\;M_{Z_L} = \frac{g}{2} v_{L} \sqrt{1+ \frac{g^2_1}{g^2+g^2_1}}, \;\; M_{Z_R} = \frac{v_{R}}{2} \sqrt{(g^2+g^2_1)} $$
Here $g_L=g_R=g$ is the $SU(2)_{L,R}$ gauge coupling whereas $g_1$ is the corresponding gauge coupling for $U(1)_{B-L}$ symmetry. Unlike in the minimal LRSM, here there is no tree level mixing between the left and right gauge bosons $W_L, W_R$. However, they can mix at one-loop level with fermions going in the loop. The mixing angle $\xi$ can be estimated as
\begin{equation}
\xi \approx \frac{\alpha}{4 \pi \sin^2{\theta_W}} \frac{m_b m_t}{M^2_{W_R}}
\end{equation}
Using $\alpha = 1/137, \sin^2{\theta_W} \approx 0.23, m_b \approx 4.2\; \text{GeV}, m_t \approx 174 \; \text{GeV}, M_{W_R} \approx 3\; \text{TeV}$, we find $\xi \approx 2 \times 10^{-7}$.

In the absence of scalar bidoublet one can not write down a Dirac mass term for fermions including quarks 
and lepton. Thus, we introduce vector-like fermions so that both left-handed and right handed fermion doublets 
of minimal left-right symmetric model can couple to each other with the following interaction Lagrangian, 
\begin{align}
\mathcal{L} & \supset Y_U (\overline{Q_L} H_L U_R+\overline{Q_R} H_R U_L) + Y_D (\overline{Q_L} H^{*}_L D_R 
+\overline{Q_R} H^{*}_R D_L) +M_U \overline{U_L} U_R+ M_D \overline{D_L} D_R\nonumber \\
& +Y_E (\overline{\ell_L} H^{*}_L E_R+\overline{\ell_R} H^{*}_R E_L) +Y_R (\overline{\ell_L} H_L N_R +\overline{\ell_R} H_R N_L) + Y_L (\ell^T_L H^*_L N_L +\ell^T_R H^*_R N_R)
\nonumber \\
& +M_E \overline{E_L} E_R+ M^D_N \overline{N_L} N_R +\frac{1}{2} M^M_N (N_L N_L + N_R N_R)+\text{h.c.}
\end{align}
After spontaneous symmetry breaking, the charge fermion mass matrices are given by
\begin{gather}
\label{2.3}
	M_{uU}    = \begin{pmatrix} 0 & Y_U v_L \\ Y^T_U v_R & M_U \end{pmatrix}, \,
	M_{dD}    = \begin{pmatrix} 0 & Y_D v_L \\ Y^T_D v_R & M_D \end{pmatrix}, \nonumber\\
	M_{e E}   = \begin{pmatrix} 0 & Y_E v_L \\ Y^T_E v_R & M_E \end{pmatrix}.
\end{gather}
The usual quarks get their Dirac masses via universal seesaw as follows,
\begin{align}
\label{2.4.0}
	M_u \approx Y^T_U \frac{1}{M_U} Y_U v_L v_R, \quad, M_d \approx Y^T_D \frac{1}{M_D} Y_D v_L v_R
\end{align}
and the mixing angles $\theta^{L,R}_U$ are found to be
\begin{align}
\label{2.4.1}
	\tan(2\theta^{}_U) \approx 2 Y^{}_U \frac{v_{L,R} M_U}{M_U^2 \pm (Y_U v_R)^2}.
\end{align}
The charged leptons get their mass as 
\begin{align}
\label{2.4.0}
	M_l \approx Y^T_E \frac{1}{M_E} Y_E v_L v_R
	\end{align}
The heavy singlet neutrino mass matrix in the basis $(N_L, N_R)$ can be block diagonalised to find the physical masses $(N_1, N_2)$ as 
$$ 
\left(\begin{array}{c}
N_L \\
N_R 
\end{array} \right) = \left(\begin{array}{cc}
		c_{\theta}  & s_{\theta} \\
   	    -s_{\theta} & c_{\theta}
\end{array} \right) \left(\begin{array}{c}
N_1 \\
N_2  
\end{array} \right) $$
where $c_{\theta} \equiv \cos{\theta}, s_{\theta} \equiv \sin{\theta}$ while the block diagonal heavy mass matrices are $M_{N_1} = M^M_N - M^D_N, M_{N_2} = M^M_N + M^D_N$. After integrating out these heavy singlet neutrinos, the light neutrino mass matrix can be written as
\begin{eqnarray}
M^{6\times 6}_\nu &=& 
	\left(\begin{array}{cc}
		M_L  & M_D \\
   	    M^T_D & M_R
\end{array} \right) 
\label{eqn:numatrix}       
\end{eqnarray}
where 
\begin{align}
\frac{M_L}{v^2_L}& = Y^T_L \left( \frac{c^2_{\theta}}{M_{N_1}} +  \frac{s^2_{\theta}}{M_{N_2}} \right) Y_L  + Y^T_R \left( \frac{s^2_{\theta}}{M_{N_1}} +  \frac{c^2_{\theta}}{M_{N_2}} \right) Y_R \nonumber \\
& +Y^T_L \left(- \frac{c_{\theta} s_{\theta}}{M_{N_1}} +  \frac{c_{\theta} s_{\theta}}{M_{N_2}} \right) Y_R  + Y^T_R \left(- \frac{c_{\theta} s_{\theta}}{M_{N_1}} +  \frac{c_{\theta} s_{\theta}}{M_{N_2}} \right) Y_L \\
&= \frac{M_D}{v_L v_R}  = \frac{M_R}{v^2_R}.
\end{align}
Assuming $v_L \ll v_R$, the light neutrino mass matrix can be written as 
\begin{eqnarray}
M_\nu = M_L- M^T_D \frac{1}{M_{R}} M_D
\end{eqnarray}
which vanishes for the above definitions of $M_L, M_D, M_R$. We therefore, forbid the terms $Y_L (\ell^T_L H^*_L N_L +\ell^T_R H^*_R N_R)$ in the Lagrangian. It is easy to see that these terms violate lepton number similar to the Majorana mass terms $M^M_N (N_L N_L + N_R N_R)$. One can introduce additional symmetries that can forbid lepton number violating Yukawa terms but allow the Majorana mass terms. For example, a $Z_4$ symmetry under which both lepton doublets and singlet leptons $N_{L,R}$ have same charges while the Higgs doublets are neutral can forbid the Yukawa terms but allow the Majorana mass terms arising dynamically from a singlet scalar. Without going into the details of such models, we study the phenomenological consequence of such a model, where only the bilinear Majorana mass terms violate lepton number by two units.

In such a case, the part of the Lagrangian relevant for neutrino mass is 
\begin{align}
\mathcal{L} & \supset Y_\nu (\overline{\ell_L} H_L N_R 
+\overline{\ell_R} H_R N_L)+ M^D_N \overline{N_L} N_R \nonumber \\
& +\frac{1}{2} M^M_N (N_L N_L + N_R N_R)+\text{h.c.}
\end{align}
The neutrino mass matrix in the basis $(\nu_L, \nu_R \equiv N)$ can have three independent terms
$$ M_L = -Y^T_\nu \frac{1}{M^M_N} Y_\nu v^2_L $$
$$ M_R = -Y^T_\nu \frac{1}{M^M_N} Y_\nu v^2_R $$
$$ M_D = Y^T_\nu \frac{1}{M^M_N} M^D_N \frac{1}{M^M_N} Y_\nu v_L v_R $$
after integrating out the heavy neutral fermions $N_{L,R}$. Considering $M^D_N = c_1 M^M_N$ (with $c_1$ being a numerical constant), the neutral lepton mass matrix in the basis $(\nu_L, \nu_R\equiv N)$ can be written as
\begin{eqnarray}
M^{6\times 6}_\nu &=& 
	\left(\begin{array}{cc}
		-Y^{T}_{\nu} \frac{1}{M^M_{N}} Y_{\nu} v^2_{L}  & c_1Y^{T}_{\nu} \frac{1}{M^M_{N}} Y_{\nu} v_{L} v_{R} \\
   	    c_1Y^{T}_{\nu} \frac{1}{M^M_{N}} Y^{}_{\nu} v_{L} v_{R} & -Y^{T}_{\nu} \frac{1}{M^M_{N}} Y_{\nu} v^2_{R}
\end{array} \right) = \left(\begin{array}{cc}
		M_L  & M_D \\
   	    M^T_D & M_R
\end{array} \right) 
\label{eqn:numatrix}       
\end{eqnarray}
In the limit $M_L \ll M_D \ll M_R$, the type-I seesaw contribution to 
the $3\times 3$ light neutrino mass is given by
\begin{eqnarray}
m_\nu = M_L- M^T_D \frac{1}{M_{R}} M_D=(1-c^2_1) M_L
\end{eqnarray}
where, in the last step, we have used the above definitions of $M_D, M_R$ to simplify
$$ M^T_D \frac{1}{M_{R}} M_D = c^2_1 Y^{T}_{\nu} \frac{1}{M^M_{N}} Y^{}_{\nu} \left (-Y^T_\nu \frac{1}{M^M_N} Y_\nu \right)^{-1} Y^{T}_{\nu} \frac{1}{M^M_{N}} Y_{\nu} v^2_L =-c^2_1Y^{T}_{\nu} \frac{1}{M^M_{N}} Y_{\nu} v^2_{L} = c^2_1M_L$$
We study the consequence of this for $0\nu \beta \beta$ in our subsequent analysis.

The vector-like fermions, crucial for the implementation of the universal seesaw mechanism are tightly constrained from direct searches. For example, the vector-like quark masses have a lower limit $m_q \geq 750-920$ GeV depending on the particular channel of decay \cite{Khachatryan:2015oba, Khachatryan:2015gza} whereas this bound gets relaxed to $m_q \geq 400$ GeV \cite{Khachatryan:2015jha} for long lived vector-like quarks. These limits however, get more uplifted by the latest analysis of the 13 TeV centre of mass energy data from the LHC. For example, the recent analysis \cite{Aaboud:2017qpr} constrains the vector-like top quark mass to be $m_T > 870-1170$ GeV depending on the weak isospin properties of it. Further constraints on vector-like quarks can be found in \cite{Aguilar-Saavedra:2013qpa}. The constraints on vector-like leptons are much weaker $m_l \geq 114-176$ GeV \cite{Aad:2015dha}. The experimental constraints put these lower bounds not only on the vector-like fermions, but also on the new gauge bosons of the model. The right handed gauge boson masses are primarily constrained from $K-\bar{K}$ mixing and direct searches at the LHC. While $K-\bar{K}$ mixing puts a constraint $M_{W_R} > 2.5$ TeV \cite{Zhang:2007fn}, direct search bounds depend on the particular channel under study. For example, the dijet resonance search in ATLAS experiment puts a bound $M_{W_R} > 2.45$ TeV at $95\%$ CL \cite{Aad:2014aqa} in the $g_L=g_R$ limit. On the other hand, the CMS search for same sign dilepton plus dijet $pp \rightarrow l^{\pm} l^{\pm} j j$ mediated by heavy right handed neutrinos at 8 TeV centre of mass energy excludes some parameter space in the $M^{\text{lightest}}_i-M_{W_R}$ plane \cite{Khachatryan:2014dka} where $M^{\text{lightest}}_i$ is the the mass of the lightest neutral fermion from right handed lepton doublets. More recently, the results on dijet searches at ATLAS experiment at 13 TeV centre of mass energy and 37 fb$^{-1}$ of $pp$ collision data have put even stronger limits on such heavy charged gauge bosons \cite{Aaboud:2017yvp}.
\section{$0\nu\beta\beta$ in LRSM with Universal Seesaw}
\label{sec2}
In minimal left-right symmetric model with universal seesaw (MLRSM-US) with additional vector-like leptons, 
the Dirac as well as Majorana masses for neutral leptons arise from Higgs fields $H_L$ and $H_R$. 
The resulting seesaw contributions to neutrino masses and their Majorana nature leads to rare process 
like neutrinoless double beta decay. The charge current interaction Lagrangian for leptons and quarks 
is given by
\begin{align}
 {\cal L}^{\rm lep}_{CC} &=
\frac{g_L}{\sqrt{2}}\left[\sum_{\alpha=e, \mu, \tau} \overline{\ell}_{\alpha}\, \gamma^\mu P_L {\nu}_{\alpha }\, W^{-}_{L\mu} + \mbox{h.c.}\right] 
                         \notag \\
&\hspace*{2cm}+\frac{g_R}{\sqrt{2}} \left[\sum_{\alpha=e, \mu, \tau} \overline{\ell}_{\alpha}\, \gamma_\mu P_R {N}_{\alpha}\, 
W^{-}_{L\mu} + {\rm h.c.}\right]\,, \notag \\
 {\cal L}^{\rm q}_{CC} &=
\left[ \frac{g_L}{\sqrt{2}} \overline{d} \gamma^\mu P_L u W_{L\mu}^- 
+\frac{g_R}{\sqrt{2}} \overline{d}\gamma^\mu P_R u W_{R\mu}^-  + {\rm h.c.}\right]\,, \notag 
\end{align}
In the present seesaw mechanism, the flavor neutrino eigenstates $\nu_\alpha \equiv \nu_{L \alpha}$ 
and $N_\beta \equiv \nu_{R \beta}$ are related to mass eigenstates $\nu_i$ and $N_i$ as,  
\begin{align}
&\nu_\alpha=U_{\alpha i} \nu_i + S_{\alpha i} N_i \nonumber \\ 
&N_\beta=T_{\beta i} \nu_i + V_{\beta i} N_i \nonumber 
\end{align}
The mixing matrices $U, V, S, T$ are given by 
\begin{equation}
\left(\begin{array}{cc}
\ U & S \\
\ T & V
\end{array}\right) = \left(\begin{array}{cc}
\ 1-\frac{1}{2}R R^{\dagger} & R \\
\ -R^{\dagger} & 1-\frac{1}{2}R^{\dagger}R 
\end{array}\right) \left(\begin{array}{cc}
\ U_L & 0 \\
\ 0 & U_R
\end{array}\right)
\end{equation}
such that $U_L, U_R$ are the diagonalising matrices of light and heavy neutrino mass matrices $M_{\nu_L}, M_{\nu_R}$ respectively. 
Here $R=M_{D} M^{-1}_{R}$. Simplifying the above equation gives rise to 
$$ U=U_L - \frac{1}{2} R R^{\dagger} U_L, \;\;\;\; S= R U_R$$
$$T = -R^{\dagger} U_L, \;\;\;\; V = U_R - \frac{1}{2} R^{\dagger}R U_R\,.$$

Within MLRSM-US with neutral leptons $\nu_\alpha$ and $N_\beta$ and with negligible $W_L-W_R$ mixing, the relevant contributions 
to neutrinoless double beta decay are as follows:
\begin{itemize}
\item  due to exchange of light left-handed and keV-MeV scale right-handed neutrinos via purely left-handed currents ($W_L-W_L$ mediation), 
\item  due to exchange of light left-handed and keV-MeV scale right-handed neutrinos via purely right-handed currents ($W_R-W_R$ mediation), 
\item  due to mixed helicity so called $\lambda$ diagrams which involves left-right neutrino mixing through mediation of $\nu_i, N_i$ neutrinos, 
\item  due to mixed helicity $\eta$ diagrams through mediation of $\nu_i, N_i$ neutrinos involving $W_L-W_R$ gauge boson mixing as well as 
       left-right neutrino mixing.
\end{itemize}

Before estimating Feynman amplitude and corresponding LNV effective Majorana mass parameters, we should have knowledge 
about the chiral structure of the matrix element with the neutrino propagator as follows,
\begin{eqnarray}
& &P_{L}\frac{\slashed{p}+m_i}{p^2-m_i^2}P_{L} \propto \frac{m_i}{p^2-m_i^2}\,  \quad\, , \quad 
   P_{R}\frac{\slashed{p}+m_i}{p^2-m_i^2}P_{R} \propto \frac{m_i}{p^2-m_i^2}\,, \nonumber \\
& &P_{L}\frac{\slashed{p}+m_i}{q^2-m_i^2}P_{R} \propto \frac{\slashed{p}}{p^2-m_i^2}\,  \quad\, , \quad 
P_{R}\frac{\slashed{p}+m_i}{p^2-m_i^2}P_{L} \propto \frac{\slashed{p}}{p^2-m_i^2}\,, \nonumber \\
\end{eqnarray}
\begin{eqnarray} \label{eq:mee}
\frac{m_i}{p^2-m^2_i} \simeq \left\{
\begin{array}{cc}
\frac{m_i}{p^2}\, ,
&  m^2_i \ll p^2\\[0.2cm]
-\frac{1}{m_i}
& m^2_i \gg p^2
\end{array} \right. 
\end{eqnarray}
and
\begin{eqnarray} \label{eq:mee}
\frac{\slashed{p}}{p^2-m^2_i} \propto \left\{
\begin{array}{cc}
\frac{1}{|p|}\, ,
&  m^2_i \ll p^2\\[0.2cm]
-\frac{|p|}{m^2_i}
& m^2_i \gg p^2\, .
\end{array} \right. 
\end{eqnarray}

\begin{figure*}[t!]
\centering
\begin{tabular}{cc}
\includegraphics[width=0.5\textwidth]{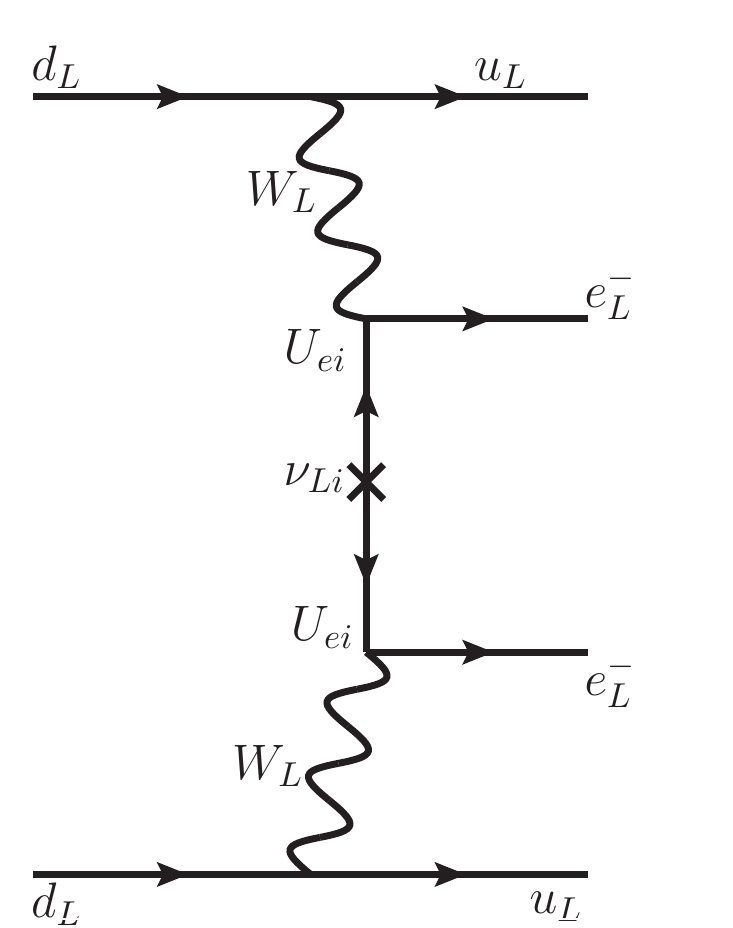} &
\includegraphics[width=0.5\textwidth]{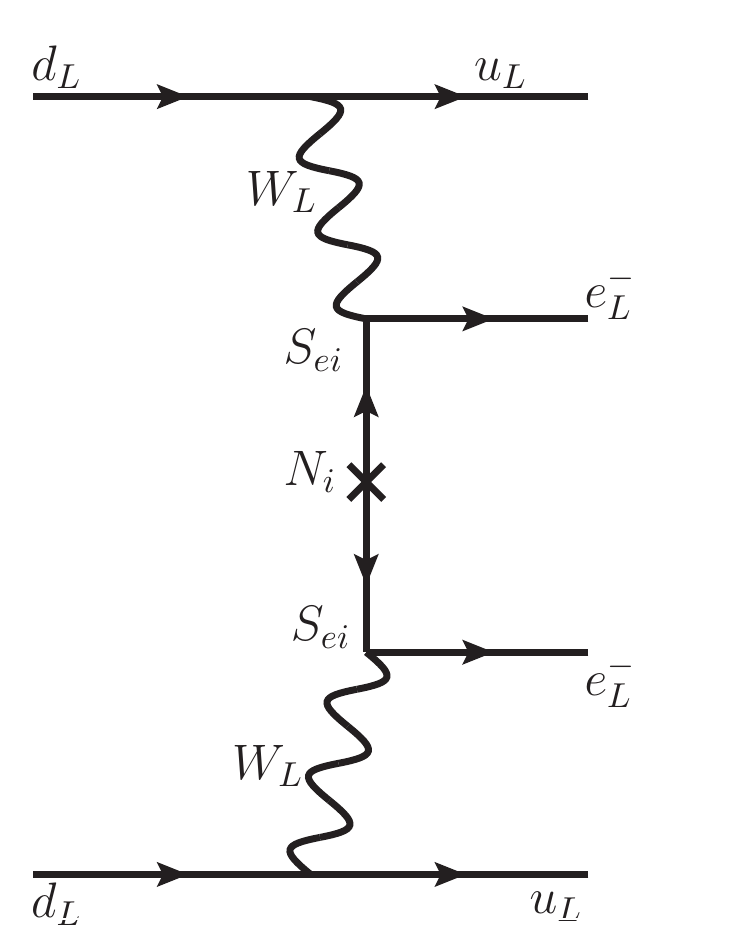}
\end{tabular}
\caption{Feynman diagram for $0\nu\beta\beta$ transition due to exchange of $\nu_i$ and $N_j$ via 
purely left-handed currents.} 
\label{fig1}
\end{figure*}
\subsection{Feynman amplitudes for different contributions to $0\nu\beta\beta$ transition}
\begin{figure*}[h!]
\centering
\begin{tabular}{cc}
\includegraphics[width=0.5\textwidth]{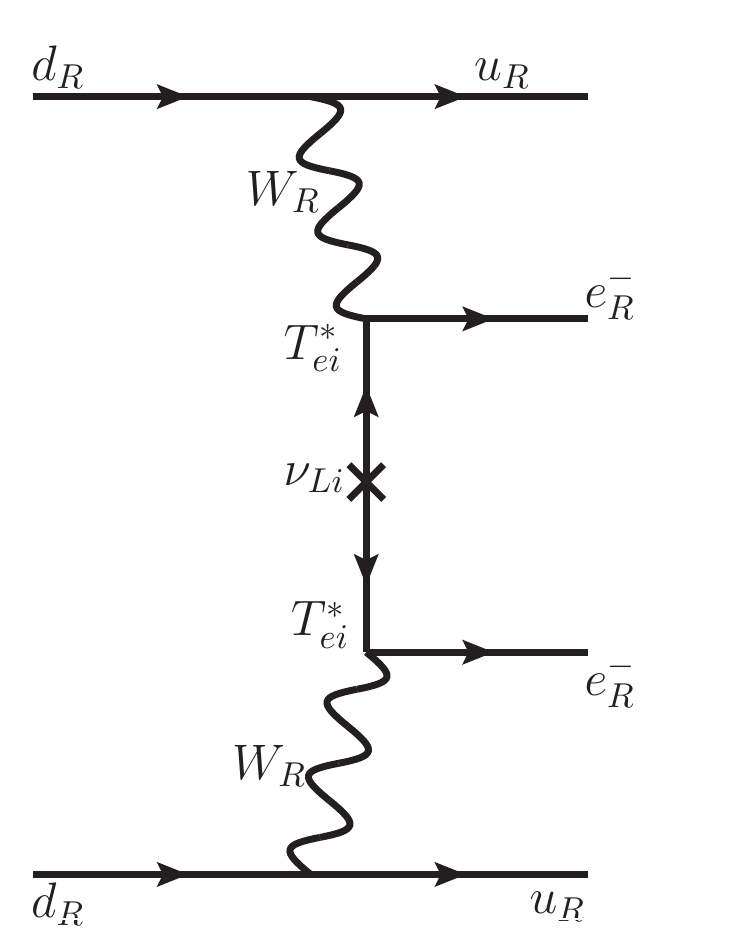} &
\includegraphics[width=0.5\textwidth]{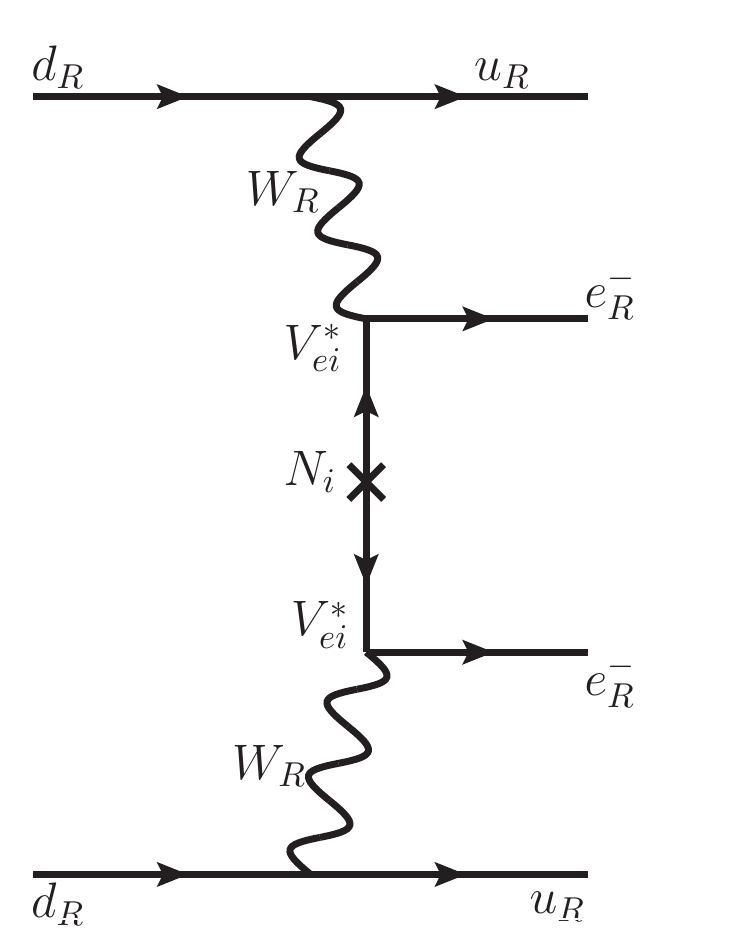}
\end{tabular}
\caption{Feynman diagram for $0\nu\beta\beta$ transition due to exchange of $\nu_i$ and $N_j$ via 
purely right-handed currents.} 
\label{fig2}
\end{figure*}

\begin{figure*}[h!]
\centering
\begin{tabular}{cc}
\includegraphics[width=0.5\textwidth]{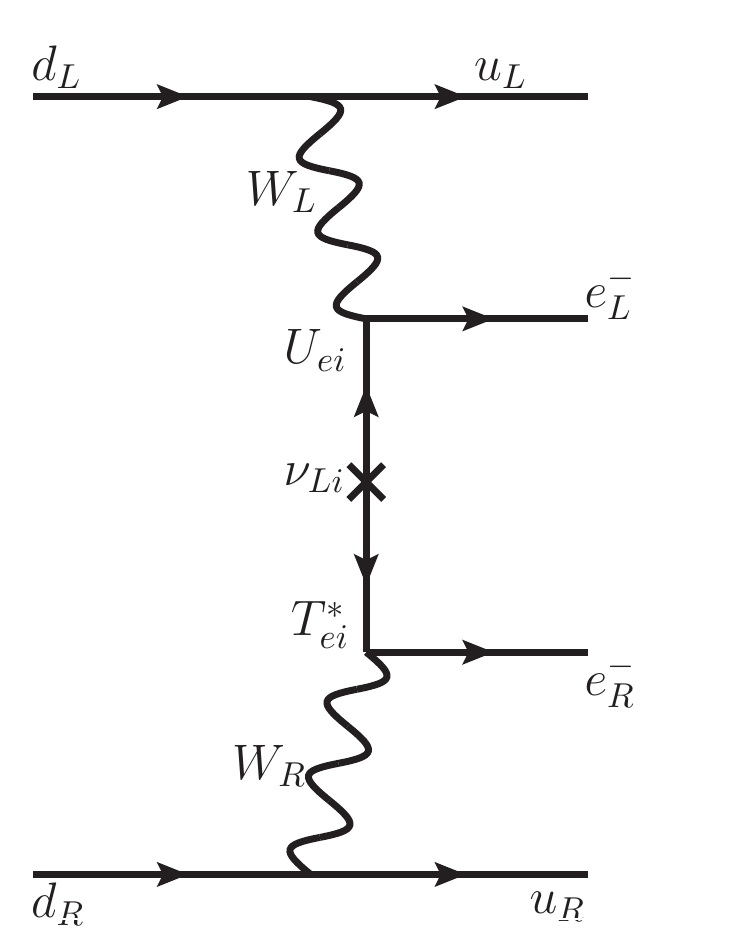} &
\includegraphics[width=0.5\textwidth]{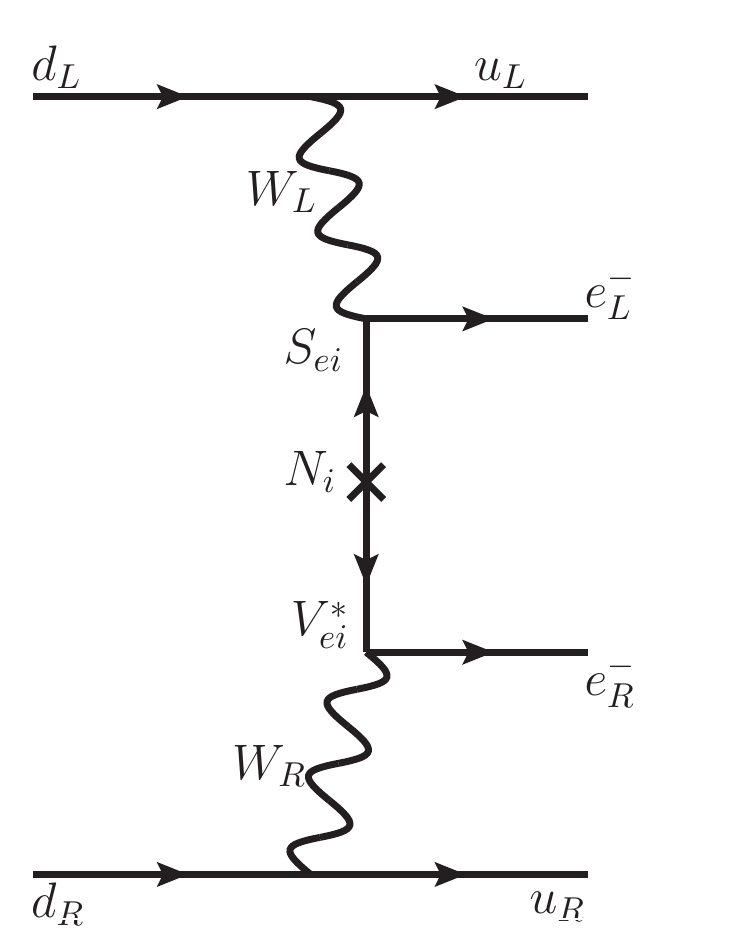}
\end{tabular}
\caption{Feynman diagram for $0\nu\beta\beta$ transition due to exchange of $\nu_i$ and $N_j$ via 
left-right neutrino mixing.} 
\label{fig3}
\end{figure*}

\begin{figure*}[h!]
\centering
\begin{tabular}{cc}
\includegraphics[width=0.5\textwidth]{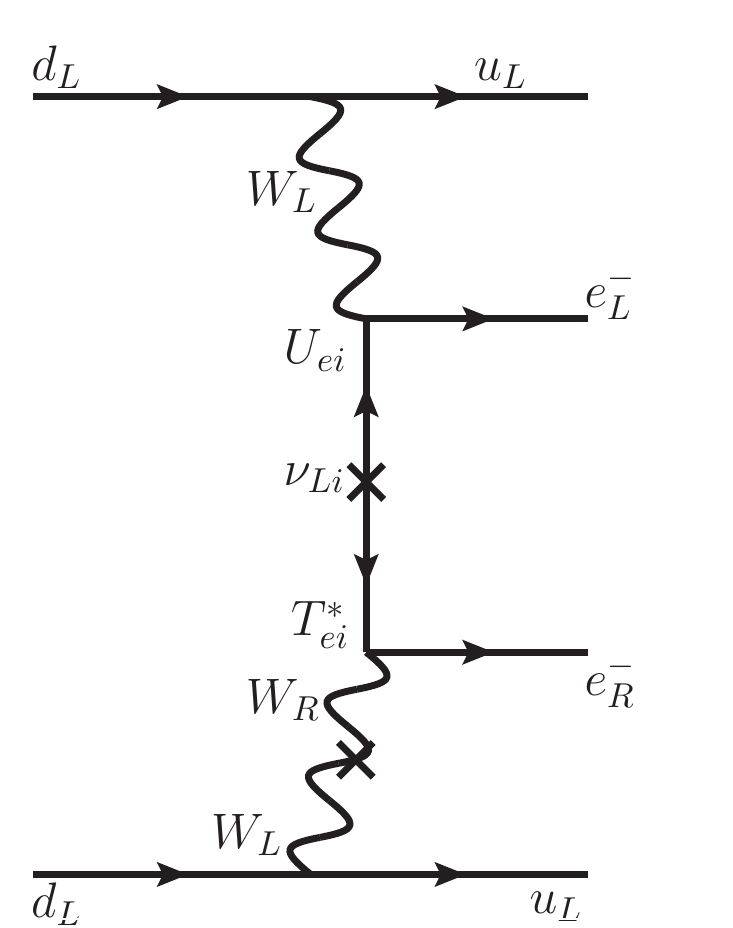} &
\includegraphics[width=0.5\textwidth]{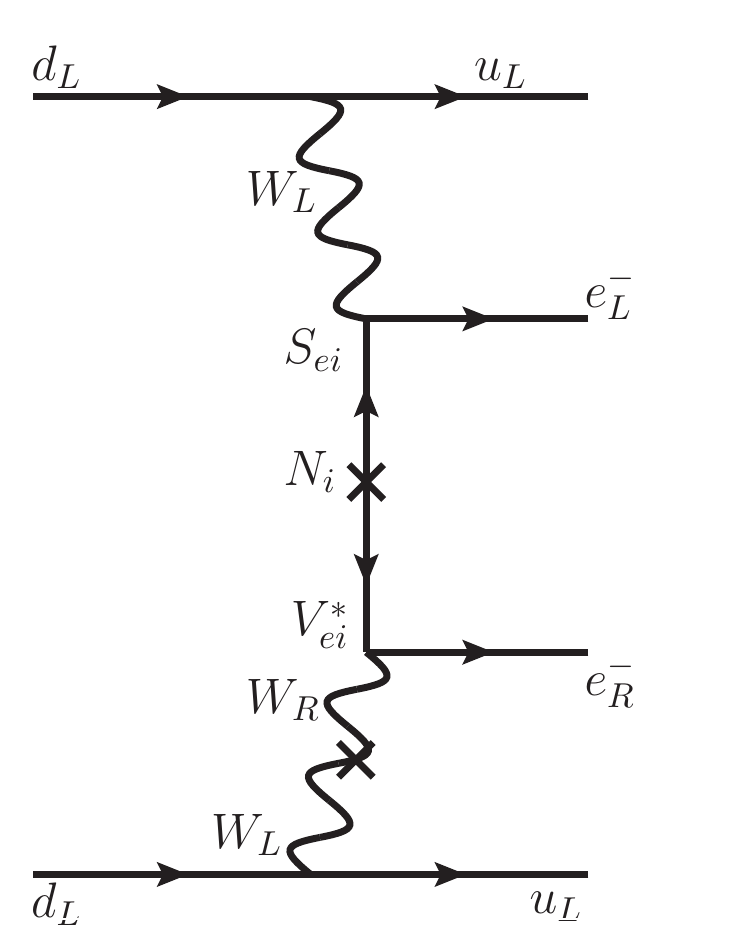}
\end{tabular}
\caption{Feynman diagram for $0\nu\beta\beta$ transition due to exchange of $\nu_i$ and $N_j$ via 
$W_L-W_R$ mixing and left-right neutrino mixing.} 
\label{fig4}
\end{figure*}

Using the notations adopted in \cite{Barry:2013xxa}, we find the amplitude for all the processes shown in Feynman diagrams of figure \ref{fig1}, \ref{fig2}, \ref{fig3}, \ref{fig4} that can contribute to $0 \nu \beta \beta$ process. The amplitudes for $0 \nu \beta \beta$ transition as shown in Fig.~\ref{fig1} due to exchange of light left-handed neutrinos and keV-MeV scale right-handed neutrinos are given by
\begin{eqnarray}
\label{eq:amp_LL} 
& &\mathcal{A}_{LL}^{\nu} \propto G^2_F \sum_{i=1,2,3} \frac{U^2_{ei}\, m_{i}}{p^2} \,, \nonumber \\
& &\mathcal{A}_{LL}^{N} \propto G^2_F \sum_{j=1,2,3} \left(\frac{S^2_{ej} M_i}{p^2} \right)\,,\nonumber                   
\end{eqnarray}
with $p \approx 100$ MeV being the average momentum exchange for the process. In the above expression, $m_i$ are the masses of light neutrinos for $i=1,2,3$ and 
$M_i$ are keV-MeV scale masses for right-handed neutrinos. 

The contribution from the left-handed neutrinos, right-handed neutrinos and $W^-_R$ exchange (Feynman diagram in Fig.~\ref{fig2}) 
can be written as 
\begin{eqnarray}
\label{eq:amp_RR} 
& &\mathcal{A}_{RR}^{\nu} \propto G^2_F \sum_{i=1,2,3} \left(\frac{M_{W_L}}{M_{W_R}} \right)^4 
            \left(\frac{g_R}{g_L} \right)^4 \frac{T^{*2}_{ei}\, m_{i}}{p^2} \,, \nonumber \\
& &\mathcal{A}_{RR}^{N} \propto G^2_F \sum_{j=1,2,3} \left(\frac{M_{W_L}}{M_{W_R}} \right)^4 
            \left(\frac{g_R}{g_L} \right)^4 \, \frac{V^{*2}_{ej}\, M_{j}}{p^2}\,, \nonumber                
\end{eqnarray}
where $M_i$ are the masses of right handed neutrinos for $i=1,2,3$ and $M_i \ll |p|$ is assumed.

The most relevant contribution from mixed helicity $\lambda$~diagram as shown in Fig.~\ref{fig3} is given by 
\begin{eqnarray}
& & \mathcal{A}^{\nu}_{\lambda} \propto {\bf G_F}^2 \left(\frac{M_{W_L}}{M_{W_R}} \right)^2 
            \left(\frac{g_R}{g_L} \right)^2 \sum_{i=1,2,3} 
U_{e\,i} T^*_{e\,i} \frac{1}{|p|} \, , \nonumber \\
& & \mathcal{A}^{N}_{\lambda} \propto {\bf G_F}^2 \sum_{j=1,2,3} \left(\frac{M_{W_L}}{M_{W_R}} \right)^2 
            \left(\frac{g_R}{g_L} \right)^2 S_{e\,j} V^*_{e\,j} \frac{1}{|p|} \, , \nonumber
\end{eqnarray}

The Feynman amplitudes for the suppressed contributions from $\eta$~diagram as displayed in Fig.~\ref{fig4} are given  by 
\begin{eqnarray}
& & \mathcal{A}^{\nu}_{\eta} \propto {\bf G_F}^2 \sum_{i=1,2,3} \left(\frac{g_R}{g_L} \right) \tan \xi
          U_{e\,i} T^*_{e\,i} \frac{1}{|p|} \, , \nonumber \\
& & \mathcal{A}^{N}_{\eta} \propto {\bf G_F}^2 \sum_{j=1,2,3} \left(\frac{g_R}{g_L} \right) \tan \xi
           S_{e\,j} V^*_{e\,j} \frac{1}{|p|}  \, , \nonumber 
\end{eqnarray}

\subsection{Effective Mass Parameters}
\begin{table}[!h]
 \centering
\vspace{10pt}
 \begin{tabular}{c|c}
 \hline \hline
Effective Mass Parameters & Analytic formula  \\
\hline
${\large \bf  m}_{\rm ee,L}^{\nu}$ & $\sum_{i=1}^3 U_{e\,i}^2\, m_{i} $  \\
${\large \bf  m}_{\rm ee,L}^{N}$   & $\sum_{i=1}^3 S_{e\,i}^2\, M_i$  \\
\hline \hline
 \end{tabular}
 \caption{Effective Majorana mass parameters from purely left-handed currents due to exchange of left-handed and 
          right-handed neutrinos. }
 \label{tab:mee_LL}
\end{table}

\begin{table}[htb!]
 \centering
\vspace{10pt}
 \begin{tabular}{c|c}
 \hline \hline
Effective Mass Parameters & Analytic formula  \\
\hline
${\large \bf  m}_{\rm ee,R}^{\nu}$ & $\left(\frac{M_{W_L}}{M_{W_R}} \right)^4 
            \left(\frac{g_R}{g_L} \right)^4\, \sum_{i=1}^3  T_{e\,i}^{*2}\, m_{i}$   \\
${\large \bf  m}_{\rm ee,R}^{N}$   & $\left(\frac{M_{W_L}}{M_{W_R}} \right)^4 
            \left(\frac{g_R}{g_L} \right)^4\, \sum_{i=1}^3 V_{e\,i}^{*2}\, M_{i}$  \\
\hline \hline
 \end{tabular}
 \caption{Effective Majorana mass parameters from purely right-handed currents due to exchange of left-handed and right-handed neutrinos. }
 \label{tab:mee_RR}
\end{table}

\begin{table}[h!]
 \centering
\vspace{10pt}
 \begin{tabular}{c|c}
 \hline \hline
Effective Mass Parameters  &  Analytic formula  \\
\hline
${\large \bf  m}_{\rm ee,\lambda}^{\nu}$   & $\left(\frac{M_{W_L}}{M_{W_R}} \right)^2 
            \left(\frac{g_R}{g_L} \right)^2\, \sum_{i=1}^3  U_{e\,i} T^*_{e\,i}\, |p|$  \\
${\large \bf  m}_{\rm ee,\lambda}^{N}$   & $\left(\frac{M_{W_L}}{M_{W_R}} \right)^2 
            \left(\frac{g_R}{g_L} \right)^2\, \sum_{j=1}^3  S_{e\,j} V^*_{e\,j}\, |p|$  \\
\hline
${\large \bf  m}_{\rm ee,\eta}^{\nu}$   & $ \left(\frac{g_R}{g_L}\right)\, \sum_{i=1}^3
 U_{e\,i} T^*_{e\,i}\, \tan \xi\, |p|$  \\
${\large \bf  m}_{\rm ee,\eta}^{N}$   & $\left(\frac{g_R}{g_L}\right)\, \sum_{j=1}^3 
 S_{e\,j} V^*_{e\,j}\, \tan \xi\, |p|$  \\
\hline \hline
 \end{tabular}
 \caption{Effective Majorana mass parameters due $\lambda$ and $\eta$ type diagrams.}
 \label{tab:mee_LR}
\end{table}
\begin{center}
\begin{table}[htb]
\begin{tabular}{|c|c|c|c|}
\hline
Isotope & $G^{0\nu}_{01} \; (\text{yr}^{-1})$  &  $\mathcal{M}^{0\nu}_\nu \equiv \mathcal{M}^{0\nu}_N$ & $\mathcal{M}^{0\nu}_{\lambda} \equiv \mathcal{M}^{0\nu}_{\eta}$ \\
\hline
$ \text{Ge}-76$ & $5.77\times10^{-15}$ & $2.58-6.64$ &$1.75-3.76$  \\
$ \text{Xe}-136$ & $3.56 \times 10^{-14}$ & $1.57-3.85$ &  $1.92-2.49$ \\
\hline
\end{tabular}
\caption{Values of phase space factor \cite{Kotila:2012zza} and nuclear matrix elements \cite{Pantis:1996py} used in the analysis}
\label{tableNME}
\end{table}
\end{center}

Combining all the contributions, one can write down the half-life of neutrinoless double beta decay as
\begin{align}
\frac{1}{T^{0\nu}_{1/2}} &= G^{0\nu}_{01} \bigg ( \lvert \mathcal{M}^{0\nu}_\nu  \eta^L_{\nu}+\mathcal{M}^{0\nu}_N \eta^L_N \rvert^2 + \lvert \mathcal{M}^{0\nu}_N \eta^R_N +\mathcal{M}^{0\nu}_\nu \eta^R_{\nu} \rvert^2 \nonumber \\
&+ \lvert \mathcal{M}^{0\nu}_{\lambda} (\eta^{\nu}_{\lambda}+\eta^{N}_{\lambda}) + \mathcal{M}^{0\nu}_{\eta} (\eta^{\nu}_{\eta}+\eta^N_{\eta}) \rvert^2 \bigg )
\label{eq:halflife}
\end{align}
where 
$$ \eta^L_{\nu} = \sum_i \frac{m_i U^2_{ei}}{m_e} =\frac{{\large \bf  m}_{\rm ee,L}^{\nu}}{m_e}, \;\;\;\;  \eta^R_{\nu}= \left ( \frac{M_{W_L}}{M_{W_R}} \right )^4  \left(\frac{g_R}{g_L} \right)^4 \sum_i \frac{m_i T^{*2}_{ei}}{m_e}=\frac{{\large \bf  m}_{\rm ee,R}^{\nu}}{m_e}$$
$$\eta^L_N=\sum_i \frac{S^2_{ei} M_i}{m_e}=\frac{{\large \bf  m}_{\rm ee,L}^{N}}{m_e},\;\;\;\; \eta^R_N= \left ( \frac{M_{W_L}}{M_{W_R}} \right )^4 \left(\frac{g_R}{g_L} \right)^4 \sum_i \frac{V^{*2}_{ei}M_i}{m_e} =\frac{{\large \bf  m}_{\rm ee,R}^{N}}{m_e}$$
$$ \eta^{\nu}_{\lambda}=\left ( \frac{M_{W_L}}{M_{W_R}} \right )^2 \left(\frac{g_R}{g_L} \right)^2 \sum_i U_{ei} T^*_{ei} =\frac{{\large \bf  m}_{\rm ee,\lambda}^{\nu}}{\lvert p \rvert}, \;\;\;\; \eta^N_{\lambda}=\left ( \frac{M_{W_L}}{M_{W_R}} \right )^2 \left(\frac{g_R}{g_L} \right)^2 \sum_i S_{ei} V^*_{ei} = \frac{{\large \bf  m}_{\rm ee,\lambda}^{N}}{\lvert p \rvert}$$
$$\eta^{\nu}_{\eta} = \left(\frac{g_R}{g_L} \right) \tan{\xi} \sum_i U_{ei} T^*_{ei}=\frac{{\large \bf  m}_{\rm ee,\eta}^{\nu}}{\lvert p \rvert}, \;\;\;\; \eta^N_{\eta} = \left(\frac{g_R}{g_L} \right) \tan{\xi} \sum_i S_{ei} V^*_{ei} = \frac{{\large \bf  m}_{\rm ee,\eta}^{N}}{\lvert p \rvert}$$
Here $m_e$ is the mass of electron. The effective mass parameters ${\large \bf  m}_{\rm ee, L, R, \lambda, \eta}^{\nu, N}$ are given in table \ref{tab:mee_LL}, \ref{tab:mee_RR}, \ref{tab:mee_LR} respectively. Also, the nuclear matrix elements involved are denoted by $\mathcal{M}$ the numerical values of which are shown in table \ref{tableNME}. The numerical values of the phase space factor $G^{0\nu}_{01}$ are also shown in the table \ref{tableNME} for different nuclei. The nuclear matrix elements for light and heavy neutrino exchanges are taken identical as all the neutrino masses lie in the range $M \ll p \sim 100$ MeV, typical momentum exchange of the process \cite{Blennow:2010th}.

\section{Lepton Flavour Violation}
\label{sec3}
\begin{figure*}[h!]
\centering
\includegraphics[width=0.75\textwidth]{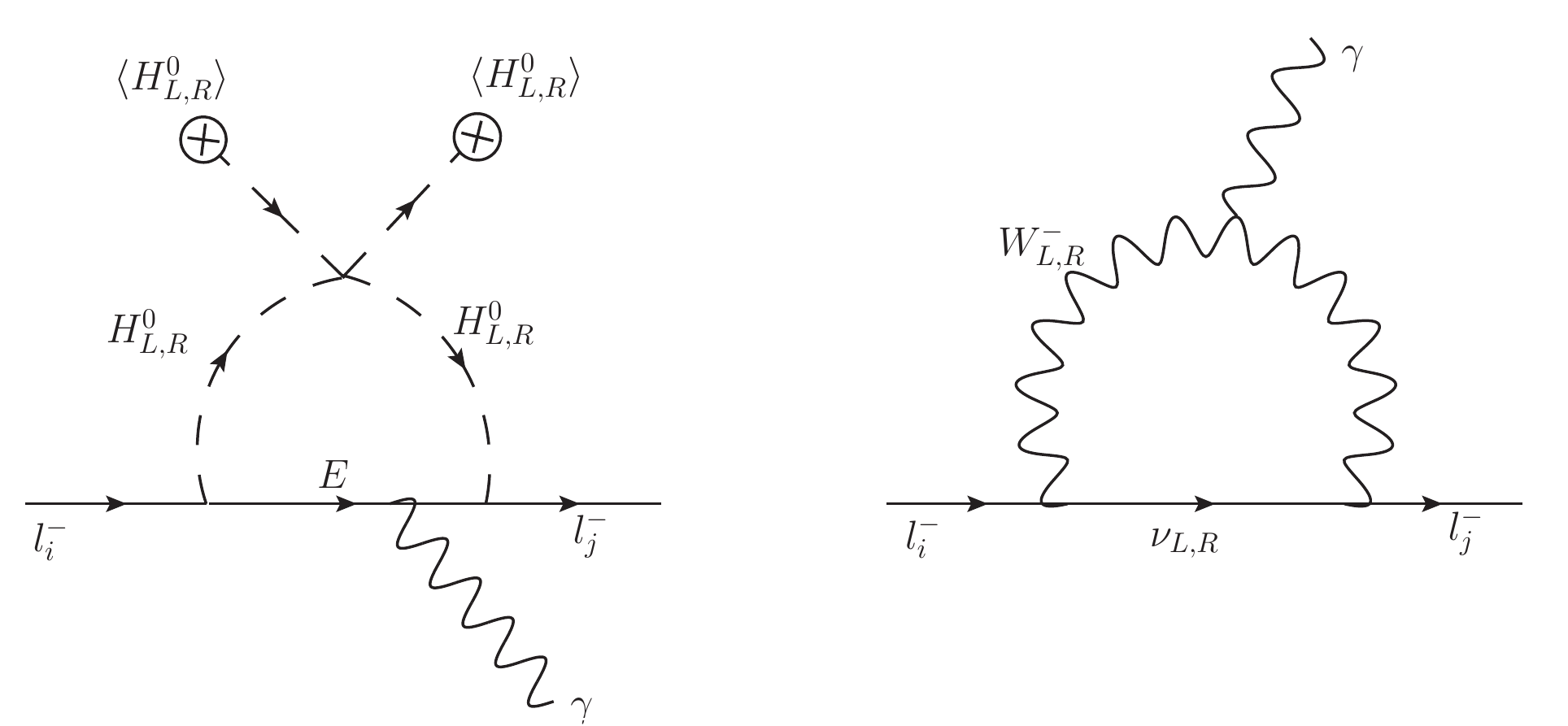}
\caption{Feynman diagram for charged lepton flavour violation} 
\label{fig5}
\end{figure*}

The new fields introduced in the model can induce LFV decays like $\mu \rightarrow e \gamma$ \footnote{For a recent review on charged lepton flavour violation, please see \cite{Lindner:2016bgg}} through one-loop diagrams with heavy charged vector-like leptons and the second scalar doublets in loop. This is shown in figure \ref{fig5}. In the SM, such LFV decays also occur at loop level but heavily suppressed due to the smallness of neutrino masses, far beyond the current experimental sensitivity \cite{TheMEG:2016wtm}. Therefore, any experimental observation of such rare decay processes will be a clear indication of BSM physics. We calculate the new physics contribution to $\Gamma (\mu \rightarrow e \gamma)$ and check for what values of new physics parameters, it can lie close to the latest bound from the MEG collaboration is $\text{BR}(\mu \rightarrow e \gamma) < 4.2 \times 10^{-13}$ at $90\%$ confidence level \cite{TheMEG:2016wtm}. 
The usual standard model contribution with $W_L, \nu_L$ in loop (shown in the right panel of figure \ref{fig5}) is given by 
\begin{equation}
\text{BR} (\mu \rightarrow e \gamma) = \frac{\sqrt{2} G^2_F m^5_{\mu}}{\Gamma_{\mu}} \lvert \sum_{i} U_{\mu i} U^*_{e i} G_{\gamma} \left ( \frac{m^2_i}{M^2_{W_L}} \right) \rvert^2 
\end{equation}
where
$$ \Gamma_{\mu}=\frac{G^2_F m^5_{\mu}}{192 \pi^3} \left( 1-8\frac{m^2_e}{m^2_{\mu}} \right)  \left( 1+ \frac{\alpha_{\text{em}}}{2\pi} (\frac{25}{4}-\pi^2) \right)$$
is the decay width of the muon and $G_{\gamma}(x)$ is the loop function given by
$$ G_{\gamma}(x) =\frac{x-6x^2+3x^3+2x^4-6x^3\log{x}}{4(1-x)^4}$$
The SM contribution is very suppressed due to the smallness of neutrino masses, of the order of $10^{-46}$. We can have three more contributions from the same diagram (right panel of figure \ref{fig3}) where the particles in the loop can be $\nu_L-W_R, \nu_R-W_R, \nu_R-W_L$. In the case of $\nu_L-W_R$, the mass of $W_L$ will be replaced by that of $W_R$ and the light neutrino mixing matrix elements $U_{\alpha i}$ will be replaced by heavy-light neutrino mixing $T_{\alpha i}$. Similarly, in $\nu_R-W_L$ case, the mass of light neutrino will be replaced by that of heavy neutrino $M_i$ and he light neutrino mixing matrix elements $U_{\alpha i}$ will be replaced by the light heavy mixing $S_{\alpha i}$. On the other hand, the diagram with $\nu_R (\equiv N)-W_R$ in the loop, we need to do the substitutions: $U \rightarrow V, m_i \rightarrow M_i, M_{W_L} \rightarrow M_{W_R}$. For all these cases, the contribution to $\text{BR} (\mu \rightarrow e \gamma)$ does not come anywhere close to the latest MEG bound. If we use the heavy neutrino mass around a keV-MeV, and $W_R$ mass at a few TeV, we can get a few order of magnitudes enhancement compared to the SM prediction, but still remains much below the experimental sensitivity. 

On the other hand, the diagram on the left panel of figure \ref{fig5} does not give any contribution as the flavour basis is identical to mass basis. This is due to the fact that the same couplings are involved in the mass of charged leptons.

\section{Numerical Analysis}
\label{sec4}
As discussed in section \ref{sec1}, we need to choose $M^D_N \neq M^M_N$ in order to have non-vanishing 
light neutrino masses. For simplicity, we consider $M^D_N$ and $M^M_N$ to be equal upto a numerical factor 
$c_1$. In this case, the light neutrino mass formula can be written as 
$$m_{\nu} = (1-c^2_1) M_L$$
This also gives rise to the same diagonalising matrices for both light and heavy neutrino mass matrices 
$U_L=U_R$. In such a case, the matrices $U, S, T, V$ can be written as 
$$ U=U_L - \frac{1}{2} R R^{\dagger} U_L=U_L-\frac{\lvert c_1 \rvert^2}{2} \left( \frac{v_L}{v_R} \right)^2 U_L  \; \text{where} \; R=M_{D} M^{-1}_{R}=c_1 \frac{v_L}{v_R},$$
$$ S=R U_R=c_1 \frac{v_L}{v_R}U_R, \;T= -R^{\dagger} U_L= -c^*_1 \frac{v_L}{v_R} U_L,$$
$$V= U_R - \frac{1}{2} R^{\dagger}R U_R=U_R - \frac{\lvert c_1 \rvert^2}{2}\left( \frac{v_L}{v_R} \right)^2 U_R.$$
Now, $U_L$ can be parametrised as the Pontecorvo-Maki-Nakagawa-Sakata (PMNS) leptonic mixing matrix
\begin{equation}
U_{\text{PMNS}} = U^{\dagger}_l U_L
\label{pmns0}
\end{equation}
if the charged lepton mass matrix is diagonal or equivalently, $U_l = \mathbb{I}$. 
The PMNS mixing matrix can be parametrised as
\begin{equation}
U_{\text{PMNS}}=\left(\begin{array}{ccc}
c_{12}c_{13}& s_{12}c_{13}& s_{13}e^{-i\delta}\\
-s_{12}c_{23}-c_{12}s_{23}s_{13}e^{i\delta}& c_{12}c_{23}-s_{12}s_{23}s_{13}e^{i\delta} & s_{23}c_{13} \\
s_{12}s_{23}-c_{12}c_{23}s_{13}e^{i\delta} & -c_{12}s_{23}-s_{12}c_{23}s_{13}e^{i\delta}& c_{23}c_{13}
\end{array}\right) U_{\text{Maj}}
\label{matrixPMNS}
\end{equation}
where $c_{ij} = \cos{\theta_{ij}}, \; s_{ij} = \sin{\theta_{ij}}$ and $\delta$ is the leptonic Dirac CP phase. 
The diagonal matrix $U_{\text{Maj}}=\text{diag}(1, e^{i\alpha}, e^{i(\beta+\delta)})$  contains the Majorana 
CP phases $\alpha, \beta$ which remain undetermined at neutrino oscillation experiments. The light neutrino 
masses that appear in the $0\nu \beta \beta$ half-life can be written in terms of the lightest neutrino mass 
and the experimentally measured mass squared differences. For normal hierarchy, the diagonal mass matrix of 
the light neutrinos can be written  as 
$$m^{\text{diag}}_{\nu} 
= \text{diag}(m_1, \sqrt{m^2_1+\Delta m_{21}^2}, \sqrt{m_1^2+\Delta m_{31}^2})$$
whereas for inverted hierarchy 
 it can be written as 
 $$m^{\text{diag}}_{\nu} = \text{diag}(\sqrt{m_3^2+\Delta m_{23}^2-\Delta m_{21}^2}, 
\sqrt{m_3^2+\Delta m_{23}^2}, m_3)$$
 The heavy neutrino masses are related to the light neutrinos as 
\begin{equation}
 M_i = \frac{v^2_R}{v^2_L} \frac{1}{1-c^2_1} m_i\, .
\end{equation}
Thus, once we choose the scale of left-right symmetry or $v_R$, we can calculate the new physics contributions 
to $0\nu\beta\beta$ by varying the lightest neutrino mass, the CP phases for different values of $c_1$. The other 
parameters like the mixing angles and mass squared differences can be varied in their $3\sigma$ global fit range 
given in table \ref{tabglobalfit}.

\begin{table}[htb]
\centering
\begin{tabular}{|c|c|c|}
\hline
Parameters & Normal Hierarchy (NH) & Inverted Hierarchy (IH) \\
\hline
$ \frac{\Delta m_{21}^2}{10^{-5} \text{eV}^2}$ & $7.03-8.09$ & $7.02-8.09 $ \\
$ \frac{|\Delta m_{31}^2|}{10^{-3} \text{eV}^2}$ & $2.407-2.643$ & $2.399-2.635 $ \\
$ \sin^2\theta_{12} $ &  $0.271-0.345 $ & $0.271-0.345 $ \\
$ \sin^2\theta_{23} $ & $0.385-0.635$ &  $0.393-0.640 $ \\
$\sin^2\theta_{13} $ & $0.01934-0.02392$ & $0.01953-0.02408 $ \\
$ \delta $ & $0-2\pi$ & $0-2\pi$ \\
\hline
\end{tabular}
\caption{Global fit $3\sigma$ values of neutrino oscillation parameters \cite{Esteban:2016qun}.}
\label{tabglobalfit}
\end{table}
\begin{figure}[h!]
\centering
\begin{tabular}{cc}
\includegraphics[width=0.5\textwidth]{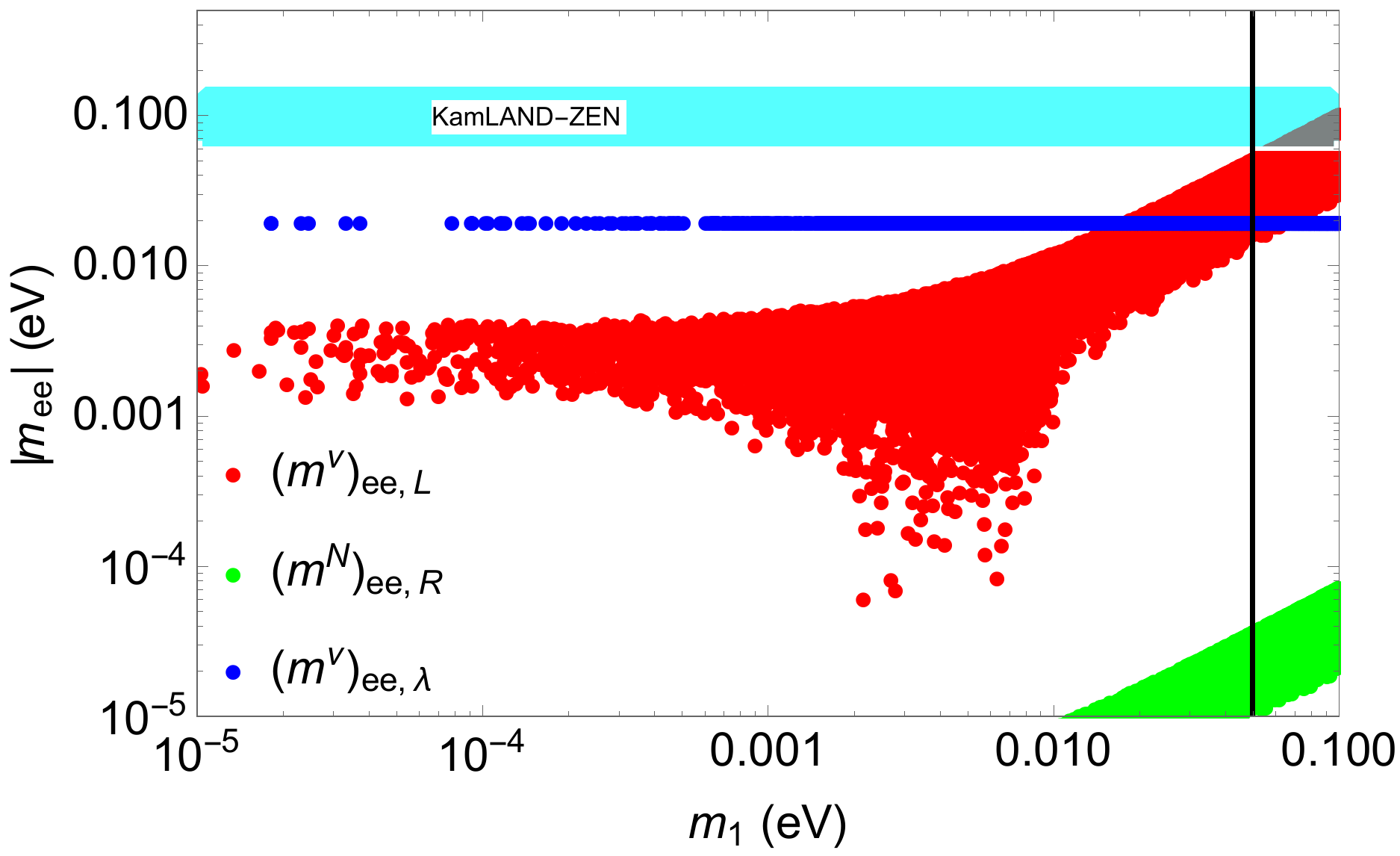}
\includegraphics[width=0.5\textwidth]{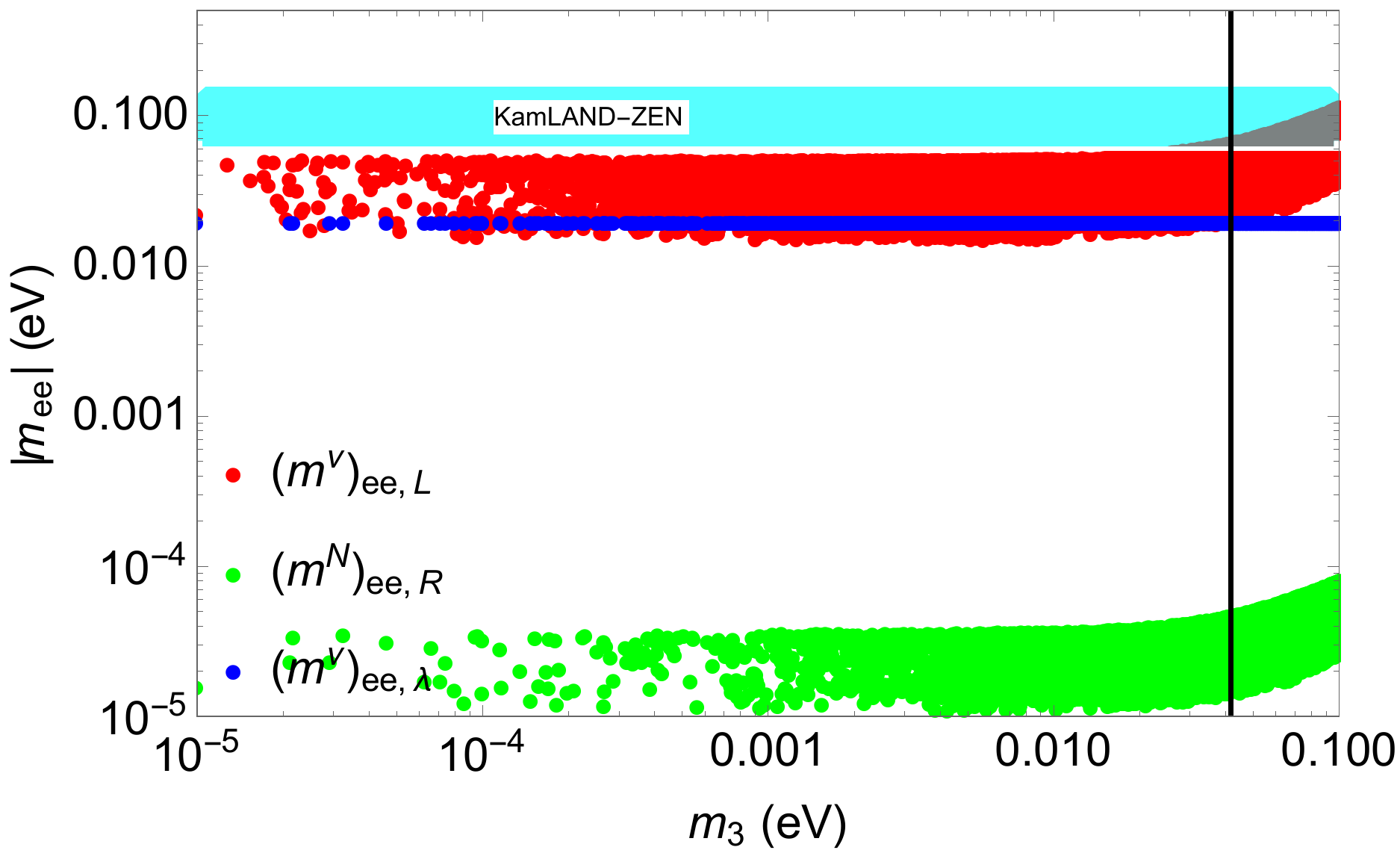}
\end{tabular}
\caption{New physics contribution to effective neutrino mass relevant for $0\nu \beta \beta$ as a function of lightest neutrino masses for $M_{W_R} = 3$ TeV, $c_1 = 10^{-5}$. The region corresponding to $m_{\rm ee} \in (61-165)$ meV set as upper bound from KamLAND-ZEN data \cite{KamLAND-Zen:2016pfg} is shaded. We also show the limit from Planck mission data as $\sum_i m_i < 0.17~$eV \cite{Ade:2015xua} on lightest neutrino mass by the vertical solid black line so that the region towards the right of this line are disfavoured.}
\label{plot:mee-0nubb-1}
\end{figure}
\begin{figure}
\centering
\begin{tabular}{cc}
\includegraphics[width=0.5\textwidth]{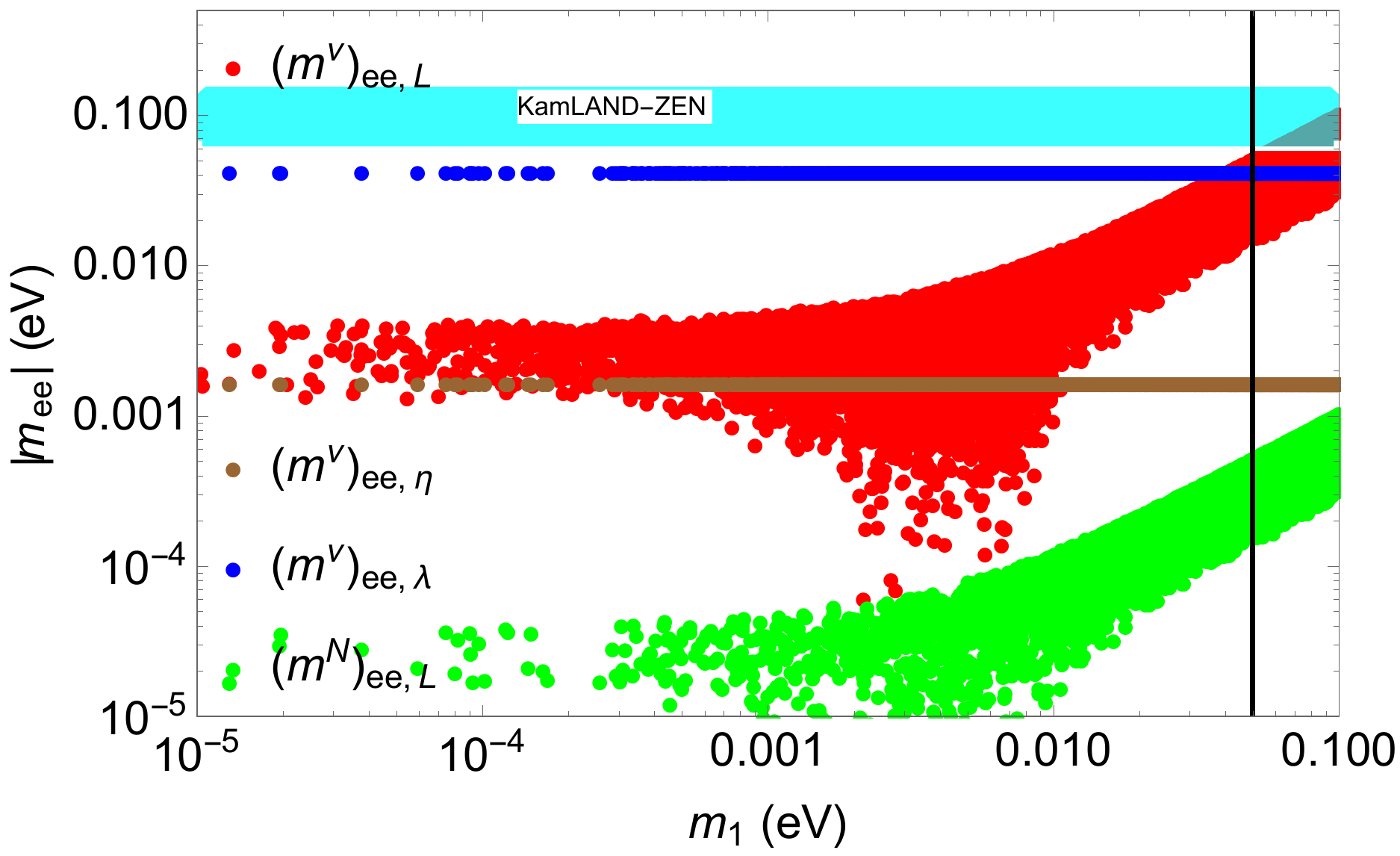}
\includegraphics[width=0.5\textwidth]{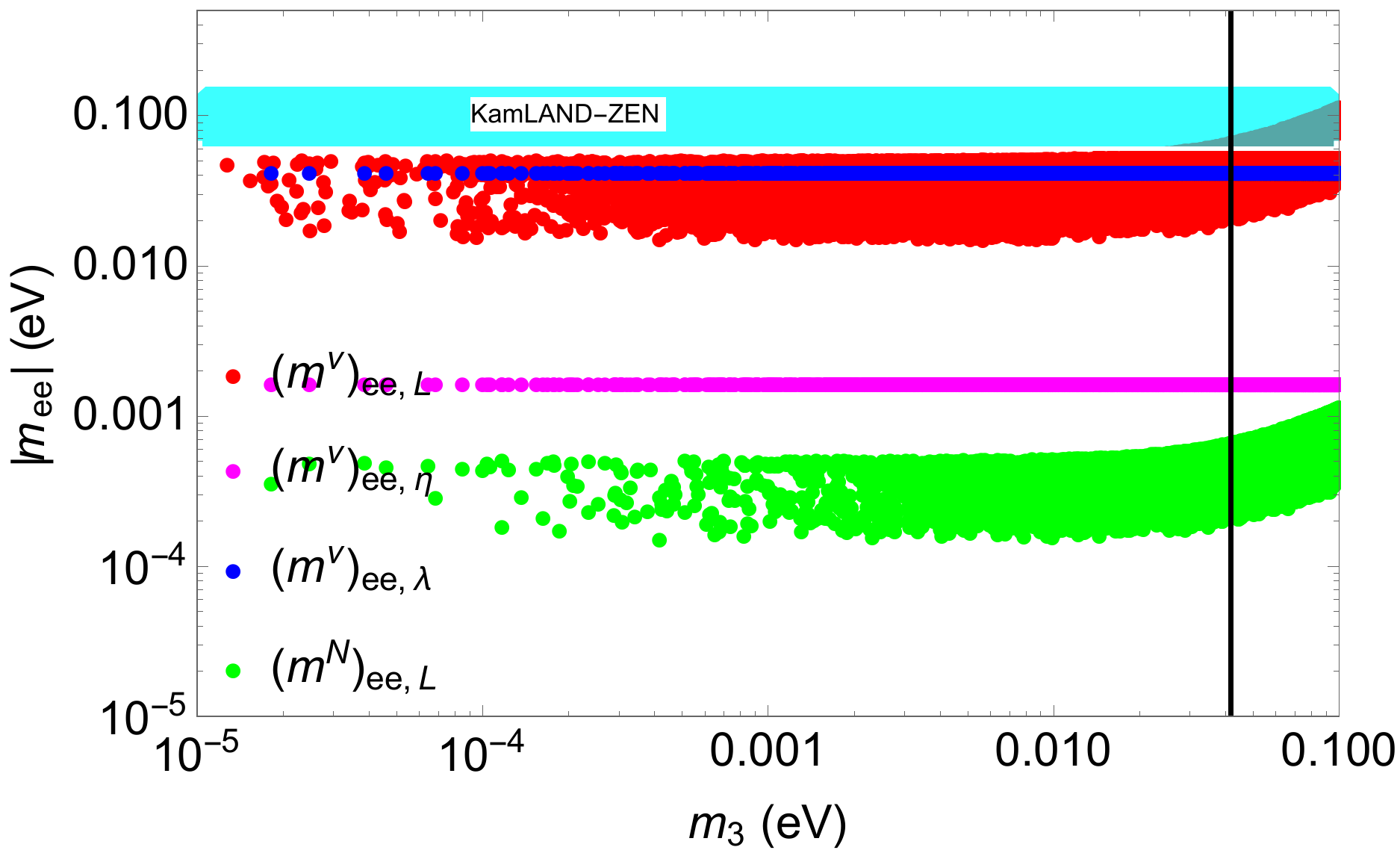}
\end{tabular}
\caption{New physics contribution to effective neutrino mass relevant for $0\nu \beta \beta$ as a function of lightest neutrino masses for $M_{W_R} = 50$ TeV, $c_1 = 0.1$. The region corresponding to $m_{\rm ee} \in (61-165)$ meV set as upper bound from KamLAND-ZEN data \cite{KamLAND-Zen:2016pfg} is shaded. We also show the limit from Planck mission data as $\sum_i m_i < 0.17~$eV \cite{Ade:2015xua} on lightest neutrino mass by the vertical solid black line so that the region towards the right of this line are disfavoured.}
\label{plot:hlife-0nubb-MWR_3TeV}
\end{figure}
\begin{figure}
\centering
\begin{tabular}{cc}
\includegraphics[width=0.75\textwidth]{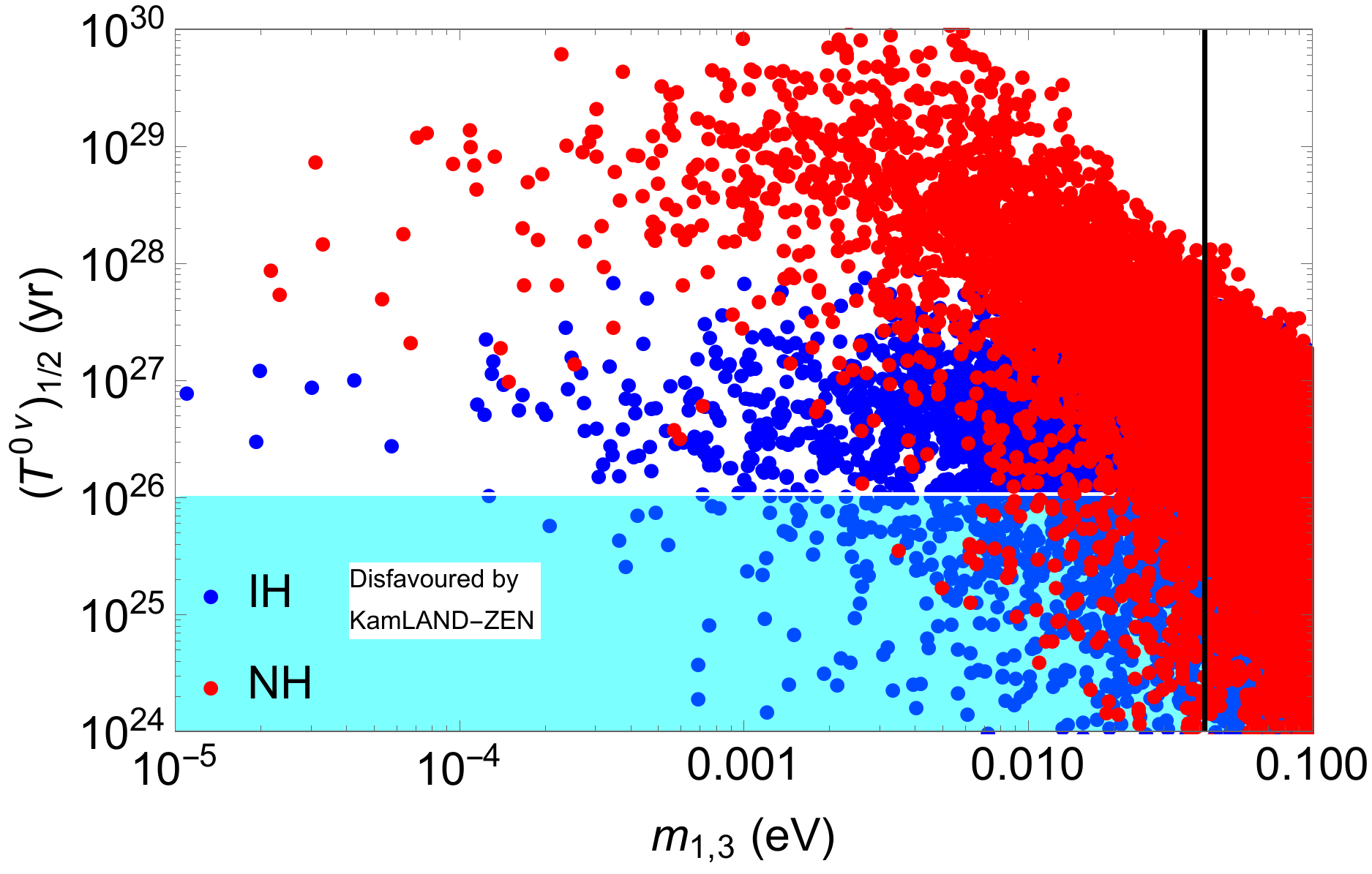}
\end{tabular}
\caption{Contribution to neutrinoless double beta decay half-life for $M_{W_R}=50$ TeV, $c_1 \in (0, 0.99)$ and $3 \sigma$ values of neutrino oscillation parameters for normal hierarchy (NH) and for inverted hierarchy (IH), respectively. The region disfavoured by KamLAND-Zen limit on half-life $T^{0\nu}_{1/2} > 1.07 \times 10^{26}$ yr \cite{KamLAND-Zen:2016pfg} is shaded. We also show the limit from Planck mission data as $\sum_i m_i < 0.17~$eV \cite{Ade:2015xua} on lightest neutrino mass by the vertical solid black line so that the region towards the right of this line are disfavoured.}
\label{plot:hlife-0nubb-MWR_100TeV}
\end{figure}

\subsection{$0\nu\beta\beta$ with $M_{W_R}=3.0$ TeV}
In the scenario where $M_{W_R} \approx 3.0~$TeV, all possible new physics contribution (apart from the usual light neutrino one in the SM) to $0\nu\beta\beta$ 
decay effective mass can be there, among which the contribution from $\lambda$ diagrams could be very large if the parameter $c_1$ is of order unity. The $\lambda-$diagram with $c_1=10^{-5}$, $|p|\simeq 100~$MeV, $M_{W_R}\simeq 3.0~$TeV and $g_L=g_R$ is estimated to be,
\begin{align}
\lvert m^{\nu}_{\rm ee, \lambda} \rvert = \left(\frac{M_{W_L}}{M_{W_R}}\right)^2 
             \bigg \lvert \sum_{i=1}^3 c_1 U_{ei} U^*_{ei} |p| \frac{v_L}{v_R}\bigg \rvert \approx \mbox{0.0189\,eV}\,.
\end{align} 
Here we ignoring a factor $(1-\frac{1}{2} \lvert c_1 \rvert^2 \left( \frac{v_L}{v_R} \right)^2)$ that appears in the definition of $U, V$ as mentioned earlier. This can clearly saturate the experimental bound from the KamLAND-Zen experiment on effective neutrino mass $m_{\rm ee} < 61-165$ meV \cite{KamLAND-Zen:2016pfg} as well as the corresponding lower limit on half-life, if $c_1$ is chosen to be one order of magnitude larger. We get similar contribution to $m^{N}_{\rm ee, \lambda}$ as well for such benchmark points. It is straightforward to see that for such small values of $c_1$ the $\eta$ diagram contributions namely $m^{\nu}_{\rm ee, \eta}, m^{N}_{\rm ee, \eta}$ remain suppressed further due to the $W_L-W_R$ mixing parameter $\xi \leq 10^{-7}$. The $W_R-W_R$ contribution also remains suppressed for this benchmark point. For example, the light neutrino mediated $W_R-W_R$ diagram contribution
\begin{align}
\lvert m^{\nu}_{\rm ee, R} \rvert \approx \left(\frac{M_{W_L}}{M_{W_R}}\right)^4 
             \bigg \lvert \sum_{i=1}^3 c^2_1 U^*_{ei} U^*_{ei} m_i \left( \frac{v_L}{v_R} \right)^2\bigg \rvert 
\end{align} 
remains suppressed compared to the light neutrino contribution $\lvert m^{\nu}_{\rm ee, L} \rvert$ approximately by an additional factor $\left(\frac{M_{W_L}}{M_{W_R}}\right)^4  c^2_1 \left( \frac{v_L}{v_R} \right)^2$ making it vanishingly small. The heavy neutrino mediated $W_R-W_R$ diagram contribution 
\begin{align}
\lvert m^{N}_{\rm ee, R} \rvert \approx \left(\frac{M_{W_L}}{M_{W_R}}\right)^4 
             \bigg \lvert \sum_{i=1}^3 \bigg [ 1-\frac{1}{2}\lvert c_1 \rvert^2 \left (\frac{v_L}{v_R} \right)^2 \bigg] U^*_{ei} U^*_{ei} M_i \bigg \rvert 
\end{align}
can be sizeable, but remains suppressed compared to the light neutrino ones for the chosen benchmark. These contributions are shown in figure \ref{plot:mee-0nubb-1} for both normal and inverted hierarchy of light neutrino masses.


\subsection{$0\nu\beta\beta$ with $M_{W_R}$ around $50$ TeV}
With $M_{W_R}\simeq 50~$TeV which corresponds to heavy neutrino masses as large as $M_{i} \approx ~$MeV, the new physics contributions to $0\nu\beta\beta$ decay effective mass could originate from both $\lambda$ and $\eta$ diagrams. While the $W_R-W_R$ mediated processes remain very much suppressed due to heavy $W_R$, one can still have sizeable $W_L-W_L$ mediated diagram mediated by heavy neutrinos. For a chosen benchmark $c_1=0.1$, the $\lambda$ and $\eta$ diagram contributions (corresponding to light neutrinos) can be estimated to be
\begin{align}
\lvert m^{\nu}_{\rm ee, \lambda} \rvert \approx \left(\frac{M_{W_L}}{M_{W_R}}\right)^2 
             \bigg \lvert \sum_{i=1}^3 c_1 U_{ei} U^*_{ei} |p| \frac{v_L}{v_R}\bigg \rvert \approx \mbox{0.04096\,eV}\,,
\end{align} 
\begin{align}
\lvert m^{\nu}_{\rm ee, \eta} \rvert \approx \bigg \lvert \sum_{i=1}^3 c_1 U_{ei} U^*_{ei} \tan{\xi} |p| \frac{v_L}{v_R}\bigg \rvert \approx \mbox{0.0016\,eV}\,.
\end{align} 
On the other hand, the heavy neutrino mediated $W_L-W_L$ diagram
\begin{align}
\lvert m^{N}_{\rm ee, R} \rvert \approx \bigg \lvert \sum_{i=1}^3 c^2_1 U^2_{ei} M_i  \left( \frac{v_L}{v_R} \right)^2 \bigg \rvert 
\end{align} 
remains suppressed by a factor $c^2_1 \left( \frac{v_L}{v_R} \right)^2 M_i/m_i= c^2_1 \frac{1}{1-c^2_1} \approx 10^{-2}$ compared to light neutrino contribution. These contributions are shown in figure \ref{plot:hlife-0nubb-MWR_3TeV} for both normal and inverted hierarchy of light neutrino masses.

We finally add all the contribution to $0\nu\beta\beta$ half-life given in equation \eqref{eq:halflife} by fixing $M_{W_R} = 50$ TeV and varying other parameters randomly. For our analysis, apart from the light neutrino oscillation parameters: two mass squared differences, three mixing angle, three phases, we have only three free parameters namely, $M_{W_R}, c_1, m_{\rm lightest} \equiv m_{1,3}$. After fixing $M_{W_R}$, we vary $c_1$ randomly in the range $(0, 0.99)$ and calculate the half-life as a function of $m_{\rm lightest} \equiv m_{1,3}$ for normal and inverted hierarchy of light neutrino masses. The result is shown in figure \ref{plot:hlife-0nubb-MWR_100TeV} in comparison to KamLAND-Zen limit on half-life $T^{0\nu}_{1/2} > 1.07 \times 10^{26}$ yr \cite{KamLAND-Zen:2016pfg} and Planck bound $\sum_i m_i < 0.17~$eV \cite{Ade:2015xua}. It is interesting to note that the new physics contribution can saturate the KamLAND-ZEN bound for lightest neutrino mass as low as $m_3 \sim 10^{-4}$ eV for inverted hierarchy even when the new physics scale or the scale of left-right symmetry is beyond the reach of any ongoing or near future collider experiments.
\begin{figure}
\centering
\begin{tabular}{cc}
\epsfig{file=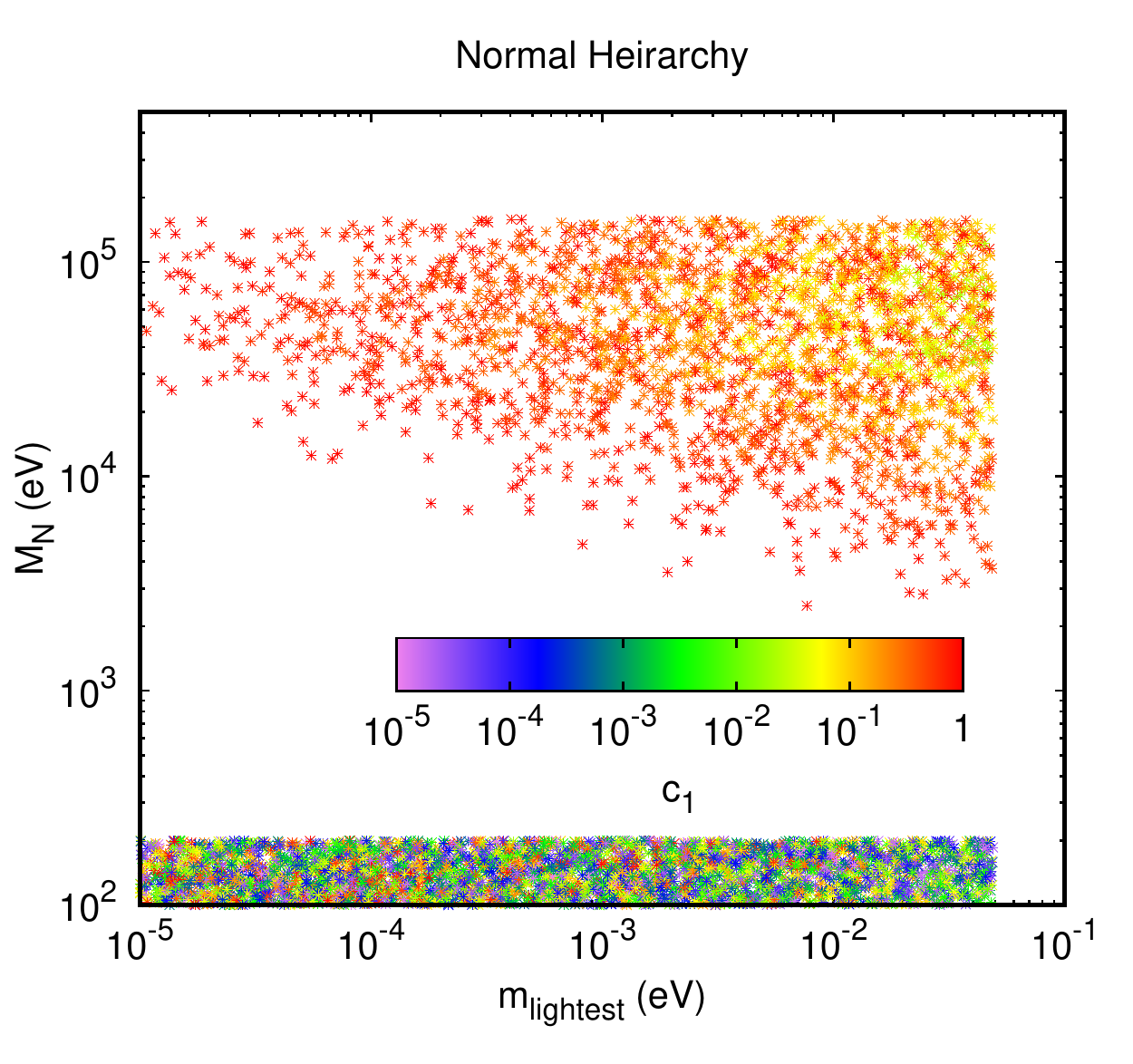,width=0.50\textwidth,clip=} &
\epsfig{file=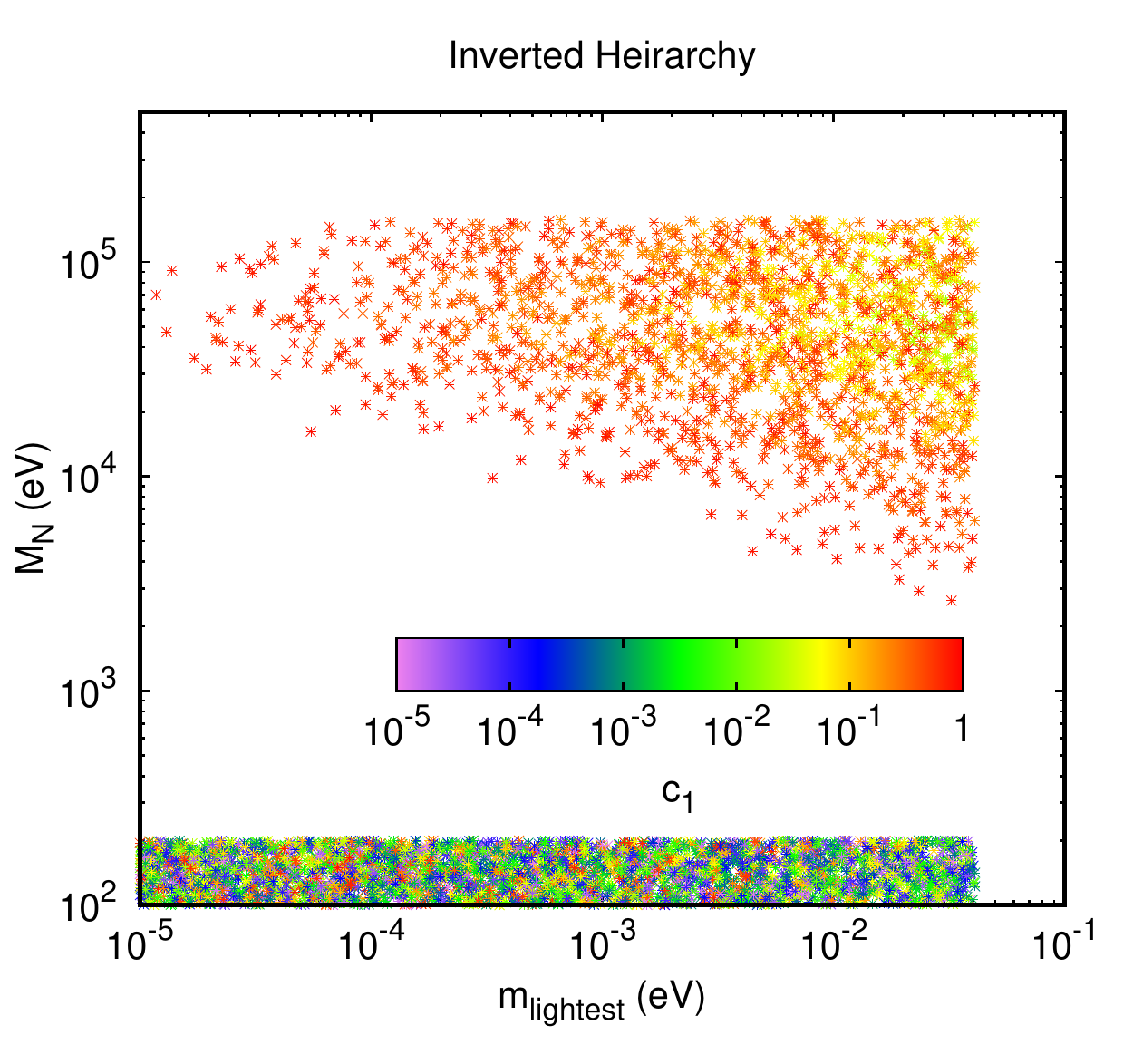,width=0.50\textwidth,clip=} \\
\epsfig{file=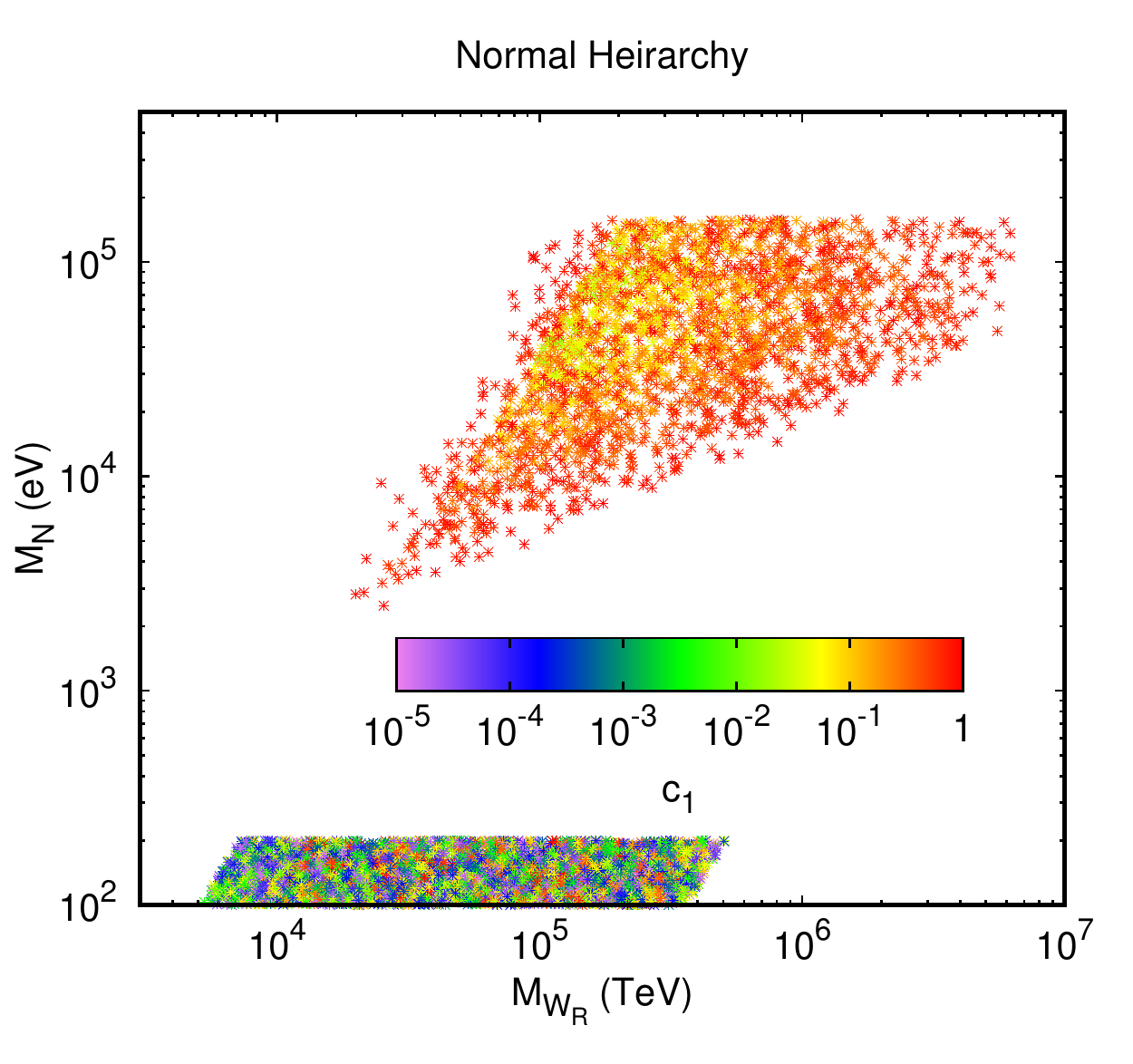,width=0.50\textwidth,clip=} &
\epsfig{file=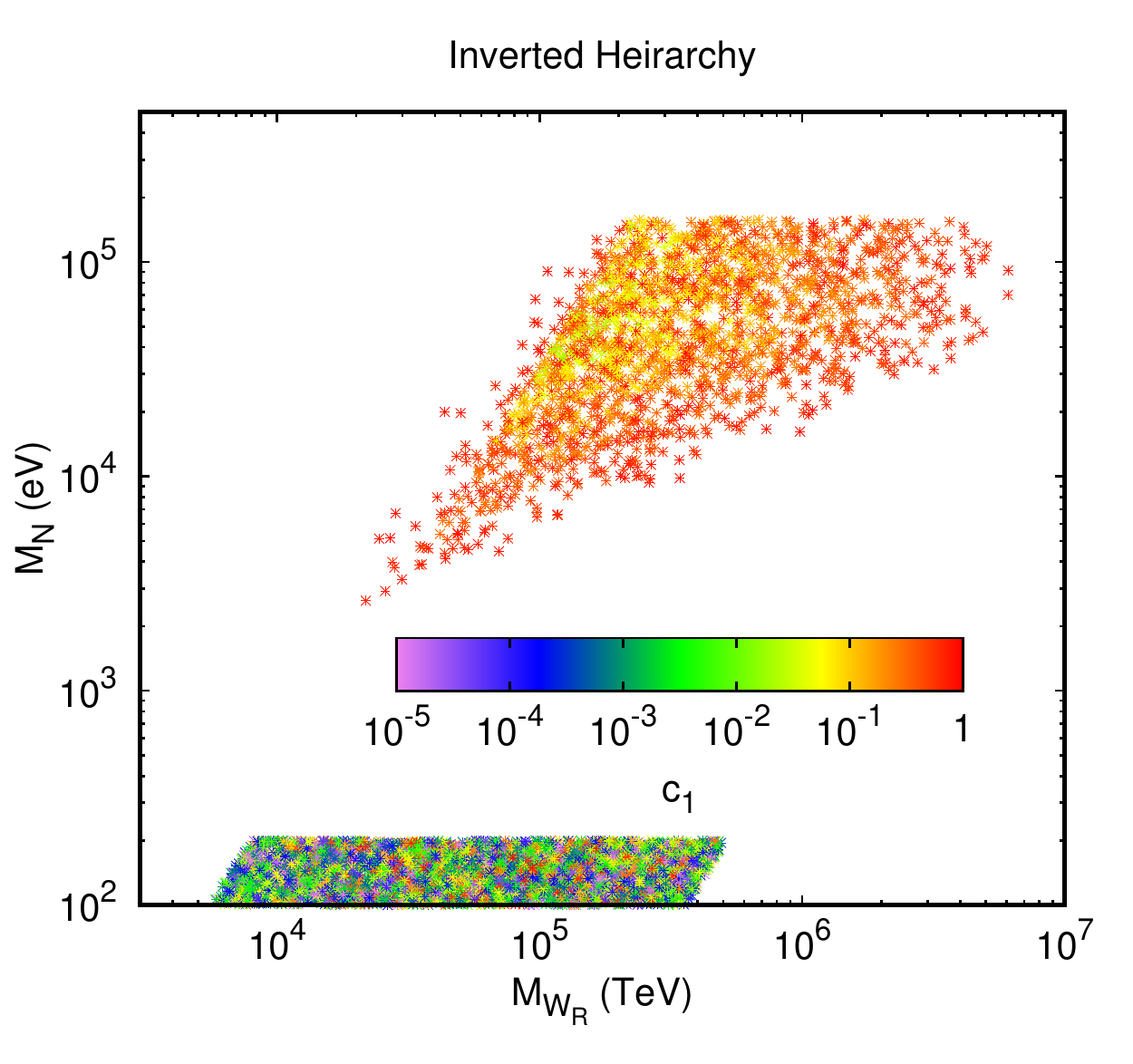,width=0.50\textwidth,clip=}
\end{tabular}
\caption{Allowed parameter space from the constraints on $0\nu\beta\beta$, generated from a random scan for one million points in the range $M_{W_R} \in (3 \; \text{TeV}, 10^7 \; \text{TeV}), \;\; c_1 \in (10^{-5}, 0.99), \;\; m_{\text{lightest}} \in (10^{-5} \; \text{eV}, \;\; 0.1 \; \text{eV})$.}
\label{scan}
\end{figure}


\subsection{Full parameter scan for $0\nu\beta\beta$}
After showing the new physics contribution to $0\nu \beta \beta$ for a few benchmark values, we perform a complete scan of parameter space for hundred thousand points and constrain the parameters from the requirement of satisfying the latest bounds from $0\nu \beta \beta$ half-life $T^{0\nu}_{1/2} > 1.07 \times 10^{26}$ yr \cite{KamLAND-Zen:2016pfg} and Planck bound $\sum_i m_i < 0.17~$eV \cite{Ade:2015xua}. We vary the light neutrino parameters in their $3\sigma$ range given in table \ref{tabglobalfit}, vary the lightest neutrino mass as $m_{\text{lightest}} \in (10^{-5} \; \text{eV}, \;\; 0.1 \; \text{eV})$ and also vary the new physics parameters randomly in the following ranges
$$ M_{W_R} \in (3 \; \text{TeV}, 10^7 \; \text{TeV}), \;\; c_1 \in (10^{-5}, 0.99).$$
The resulting parameter space is shown for both the hierarchies of light neutrino masses in figure \ref{scan}. It is interesting to see that the parameter $c_1$ is allowed to take almost any value in the chosen range, provided the values for $M_{W_R}, M_N$ are correctly chosen. This is in contrast with what we found in the analysis for effective neutrino mass for different new physics contribution. In that analysis, while referring to figure \ref{plot:mee-0nubb-1}, we concluded that for $M_{W_R} = 3$ TeV, the parameter $c_1$ has to be kept very small $(\leq 10^{-5})$ in order to keep the $\lambda$ diagram contribution within allowed range. However, as the contribution to half-life $T^{0\nu}_{1/2}$ comes from sum over all possible contributions, as shown in equation \eqref{eq:halflife}, such large contributions from individual diagrams can interfere destructively with others in order to keep the total decay rate or half-life within experimental limits. Although the values of $c_1$ are quite homogeneously distributed across the bands in figure \ref{scan}, the scattered points at the uppermost parts of these bands correspond to $c_1$ values very close to $0.99$, the upper limit chosen for the scan. This is because only such values of $c_1$ corresponds to very large values of right handed neutrino mass for a given $M_{W_R}$ due to the additional enhancement coming from the $1/(1-c^2_1)$ factor in the definition of right handed neutrino mass, discussed before. The gap between the high mass and low mass regimes of right handed neutrinos arise due to the supernova bounds which we discuss in the next section.


\section{Constraints from Cosmology and Supernova}
\label{sec4a}
Since there exists the possibility of very light right handed neutrinos in the keV scale, such a scenario can be constrained from cosmology as well as astrophysical observations. Cosmological implications for sterile neutrinos with masses from eV scale to GeV-TeV scale are well described in the recent review \cite{whitekev}. Here we focus only on the eV-keV range as the above analysis was primarily focussed on generating sterile neutrino masses in this range. Due to gauge interactions, such keV neutrinos can be thermally produced in the early Universe and can contribute to the effective relativistic degrees of freedom. Such extra relativistic degrees of freedom may affect the big bang nucleosynthesis (BBN) predictions as well as cause changes in the cosmic microwave background (CMB) spectrum. Planck 2015 data put constraints on this as $N_{\text{eff}}=3.15 \pm 0.23$ \cite{Ade:2015xua} while the 2018 data from Planck collaboration makes this bound stronger  $N_{\text{eff}}=2.99 \pm 0.17$ \cite{Ade:2018xua} at $68\%$ CL. Using these bounds, one can constrain the gauge interactions of the light sterile neutrinos. The gauge interactions for such a light keV neutrino can be written as (shown earlier)
\begin{align}
 {\cal L}^{\rm lep}_{CC} &=
\frac{g_L}{\sqrt{2}}\left[\sum_{\alpha=e, \mu, \tau} \overline{\ell}_{\alpha}\, \gamma^\mu P_L {\nu}_{\alpha }\, W^{-}_{L\mu} + \mbox{h.c.}\right] 
                         \notag \\
&\hspace*{2cm}+\frac{g_R}{\sqrt{2}} \left[\sum_{\alpha=e, \mu, \tau} \overline{\ell}_{\alpha}\, \gamma_\mu P_R {N}_{\alpha}\, 
W^{-}_{L\mu} + {\rm h.c.}\right]\,, \notag \\
 {\cal L}^{\rm lep}_{NC} &=
 \frac{e}{2 \sin{\theta_W} \cos{\theta_W}} \left[ J^{\mu}_1 Z_{L\mu} + J^{\mu}_2 Z_{R \mu} \right]\,, \notag 
\end{align}
where (ignoring left-right gauge mixing)
$$J^{\mu}_1 = \sum_{\alpha=e, \mu, \tau} (\overline{\nu_{\alpha}} \; \overline{N_{\alpha}}) \gamma^{\mu} \left[A^1_L P_L+A^1_R P_R \right] (\nu_{\alpha} \; N_{\alpha})^T, $$
$$ J^{\mu}_2 = \sum_{\alpha=e, \mu, \tau} (\overline{\nu_{\alpha}} \; \overline{N_{\alpha}}) \gamma^{\mu} \left[A^2_L P_L+A^2_R P_R \right] (\nu_{\alpha} \; N_{\alpha})^T, $$
$$ A^1_L = 2T^L_{3\nu}, A^1_R = 0, A^2_L = \frac{2\sin^2{\theta_W}}{\sqrt{\cos{2\theta_W}}} T^L_{3\nu}, A^2_R=\frac{2\cos^2{\theta_W}}{\sqrt{\cos{2\theta_W}}} T^R_{3N}. $$
Since light active neutrinos decouple from the rest of the Universe around the BBN epoch $(T^D_{\nu} \sim \mathcal{O}(\rm MeV))$, light right handed neutrinos can possibly evade BBN bounds if they decouple much earlier so that $T^D_{N} > T^D_{\nu}$. The $N$ contribution to $N_{\text{eff}}$ gets diluted due to the decrease in effective relativistic degrees of freedom $g_*$ as the Universe cools down from $T^D_{N}$ to $T^D_{\nu}$:
$$ N_{\text{eff}} \approx 3 + 3 \left (\frac{g_*(T^D_{\nu})}{g_*(T^D_{N})} \right )^{\frac{4}{3}} $$
Since $g_*(T^D_{\nu}) = 10.75$ for the relativistic degrees of freedom in SM at $T=T^D_{\nu} \approx 1$ MeV, the right handed neutrinos can evade the Planck 2018 bound if $g_*(T^D_{\nu_R}) > 125$, which is possible only when the corresponding decoupling temperature is more than the electroweak scale. The decoupling temperature of right handed neutrinos can be calculated by following the same procedure for left handed neutrinos and replacing the $W_L$ mass with $W_R$. In terms of $T^D_{\nu}$, it can be written as
$$ T^D_{N} \approx (g_*(T^D_{N})^{1/6} \left ( \frac{M_{W_R}}{M_{W_L}} \right )^{4/3}  T^D_{\nu} $$
Demanding $T^D_{N} > 200$ GeV and taking $T^D_{\nu} \approx 1$ MeV, we can arrive at the bound on $M_{W_R}$ as 
\begin{equation}
M_{W_R} > 418 \; \text{TeV}
\end{equation}
Please note that such bounds can be made weaker by including higher CL bounds on $N_{\text{eff}}$  (as well as different datasets) as shown in \cite{Ade:2015xua, Ade:2018xua}. For example, taking the upper bound as $N_{\text{eff}} < 3.7$, gives a bound on $W_R$ gauge boson as 
\begin{equation}
M_{W_R} > 4.3 \; \text{TeV}
\end{equation}
which lies closer to the reach of collider experiments.

Apart from cosmology, we can constrain such light sterile neutrinos from supernova data as well. If a sterile or right handed neutrino is produced inside the core of a star, it can escape the core resulting in rapid cooling which can be in contradiction with observations. Such a sterile neutrino can be produced either directly in scattering or through active-sterile oscillations. Direct production of sterile neutrino through scattering $\sigma (\nu X \rightarrow N X) = A \sigma (\nu X \rightarrow \nu X)$ (where $X \equiv e, p, n$) is constrained by such observations and rules out the parameter $A$ in the range $10^{-4} \lesssim A \lesssim 10^{-1}$ \cite{SN1}. In the model such scattering can happen wither through active-sterile mixing or through left-right gauge mixing. Since the left-right gauge mixing is already very small, as mentioned earlier, we use this bound to constrain the active-sterile mixing. The same active-sterile mixing is also constrained by using the bounds on sterile neutrino production from oscillations inside the core \cite{SN2, SN3}. Such bounds rule out the active-sterile neutrino mixing in the range \cite{SN2}
$$ 7 \times 10^{-10} \lesssim \sin^2{\theta_{\nu N}} \lesssim 2 \times 10^{-2} \; \text{for} \; \Delta m^2_{N\nu} \gtrsim 2 \times 10^9 \; \text{eV}^2, $$
$$ 5 \times 10^{4} \; \text{eV}^2 \lesssim \lvert \Delta m^2_{N\nu}  \sin{2\theta_{\nu N}} \rvert \lesssim 3 \times 10^{8} \; \text{eV}^2 \; \text{for} \; \Delta m^2_{N\nu} \lesssim 2 \times 10^9 \; \text{eV}^2. $$
Here $\theta_{\nu N}, \Delta m^2_{N\nu}$ are active-sterile mixing angle and mass squared difference respectively. As pointed out by \cite{SN2}, these bounds do not apply if all the right handed neutrinos have masses below 200 eV. We use these constraints to find the allowed parameter space of the model. The resulting parameter space from all the constraints are shown in figure \ref{scan}. As pointed out before, the gap between scattered points in high mass regime of right handed neutrinos and that in the low mass regime arises due to the supernova bounds that apply differently in high and low mass regime. The low mass regime for right handed neutrinos in figure \ref{scan} arises as the supernova bounds do not apply there. 
\section{Summary and Conclusion}
\label{sec5}
We have studied the new physics contribution to neutrinoless double beta decay and charged lepton flavour violation in a minimal left-right symmetric model where all the fermions acquire masses through a universal seesaw mechanism. Due to different ways of generating fermion mass and breaking the left-right gauge symmetry all the way down to the SM one, the leading contributions to $0\nu \beta \beta$ and LFV decay are very different in this model, in comparison to the usual LRSM with type I and type II seesaw for neutrino masses. One interesting feature of this model is the presence of light right handed neutrinos in the keV-MeV range, even if the scale of the theory is as high as $10^7$ TeV, by virtue of a universal seesaw framework. Since the heavy neutrinos have masses lighter than typical momentum exchange of $0\nu \beta \beta$ process, their contributions to $0\nu \beta \beta$ becomes different compared to the usual LRSM with much heavier right handed neutrinos. We identify all possible diagrams contributing to $0\nu\beta \beta$ in this model and find that the dominant individual contribution can also come from the diagrams with light-heavy neutrino mixing apart from the purely left and purely right handed diagrams. However, when we combine all the contributions, the heavy-light mixing contribution vanishes due to unitarity of the total mass matrix. For certain choices of parameters, the model predictions can saturate the current experimental bounds on $0\nu \beta \beta$. This offers a very interesting probe of such high scale LRSM which may not be directly accessible at collider experiments, due to poor sensitivity to very light right handed neutrinos. We also check the possible sources of charged lepton flavour violating decay $\mu \rightarrow e \gamma$ which is tightly constrained from present experimental data. We find that the charged gauge boson plus neutral fermion mediated diagrams to this process remain suppressed for all region of parameter space. We also derive the bounds from BBN, CMB constraints on effective relativistic degrees of freedom and shown that the model parameters can be consistent with them (even for TeV scale $W_R$ and sub-MeV right handed neutrinos) if we apply the least conservative bounds on $N_{\rm eff}$. The supernova bounds can however be very strict and can rule out some part of the existing parameter space, as shown in figure \ref{scan}.

The model also can have very interesting cosmological signatures due to the existence of light right handed neutrinos. For example, a keV scale right handed neutrino could be long lived enough to play the role of warm dark matter \cite{Bezrukov:2009th, Nemevsek:2012cd}. However, such a long-lived keV neutrino is usually overproduced as shown in these works and require additional mechanism like entropy dilution \cite{entropy2} to satisfy the relic. Although there exists several other particles like heavier right handed neutrinos that can play the role of diluters, the requirement of correct dilution constrains the masses and couplings of such diluters severely \cite{Bezrukov:2009th, Nemevsek:2012cd} and we leave such combined analysis to future works.

Since the right and left handed neutrino masses are proportional to each other, such a eV scale right handed neutrino may arise if the lightest left handed neutrino mass is vanishingly small. Such a situation will not only be interesting for neutrino oscillation experiments like MiniBooNE \cite{Aguilar-Arevalo:2018gpe}, but will also constrain the $W_R$ mass in order to satisfy the Planck bound on $N_{\text{eff}}$ \cite{Borah:2016lrl, Borah:2017leo}. If the right handed neutrinos are in the sub GeV regime, they can also play a role in creating baryon asymmetry through neutrino oscillations \cite{Canetti:2012kh}. We leave such studies in this particular model for future works.

\acknowledgments
The authors would like to thank the organisers of the discussion meeting \textit{Candles of Darkness} (ICTS/Prog-candark/2017/06) at the International Centre for Theoretical Sciences, Bangalore, India during June 5-9, 2017 for support and hospitality where part of this work was completed.
\providecommand{\href}[2]{#2}\begingroup\raggedright\endgroup

\end{document}